\documentclass[superscriptaddress, aps,pre]{revtex4-1}
\makeatletter
\newcommand*{\rom}[1]{\expandafter\@slowromancap\romannumeral #1@}
\makeatother
\usepackage{color}
\usepackage{graphicx}
\usepackage{hyperref}
\usepackage{amsmath,amssymb}
\usepackage{epstopdf}
\usepackage{subcaption}
\usepackage{float}
\graphicspath{{figs/}}
\newcommand\Tstrut{\rule{0pt}{2.6ex}}       
\newcommand\Bstrut{\rule[-0.9ex]{0pt}{0pt}} 
\newcommand{\TBstrut}{\Tstrut\Bstrut} 
\def\beq{\begin{equation}}
\def\eeq{\end{equation}}
\def\bea{\begin{eqnarray}}
\def\eea{\end{eqnarray}}

\begin{document}

\title{Heat capacities and thermodynamic geometry in deformed Jackiw–Teitelboim gravity}
\author {H. Babaei-Aghbolagh}
\email{h.babaei@uma.ac.ir}
\affiliation{Department of Physics, University of Mohaghegh Ardabili, P.O. Box 179, Ardabil, Iran}
\author {Mirmani Mirjalali}
\email{Mirmanimirjalali4@gmail.com}
\affiliation{Faculty of Physics, Shahrood University of Technology, P.O. Box 3619995161, Shahrood,
	Iran}
\author{Davood Mahdavian Yekta}
\email{d.mahdavian@hsu.ac.ir}
\affiliation{Department of Physics, Hakim Sabzevari University, P.O. Box 397, Sabzevar, Iran}
\author {Komeil Babaei Velni}
\email{babaeivelni@guilan.ac.ir}
\affiliation{Department of Physics, University of Guilan, P.O. Box 41335-1914, Rasht, Iran}\author {Hosein Mohammadzadeh}
\email{mohammadzadeh@uma.ac.ir}
\affiliation{Department of Physics, University of Mohaghegh Ardabili, P.O. Box 179, Ardabil, Iran}


\begin{abstract}

We study the thermodynamics of charged AdS black holes in deformed Jackiw-Teitelboim (dJT) gravity and their phase structures. In this regard, we will find some critical values for the temperature, entropy and charge of the corresponding black holes. We also compute the heat capacities, expansion coefficient and isothermal compressibility as thermodynamic response functions and study their behaviors at the critical points. It will be shown that these variables satisfy the Ehrenfest's equations in the case of second-order phase transition. We employ different formalisms to investigate thermodynamic geometry, such as Weinhold, Ruppeiner and new thermodynamic geometry, then analyze the singularities of the thermodynamic curvatures in this context. We show that these singularities are also correspond to the divergences of the response functions which indicating the critical points of phase transitions.
\end{abstract}
\maketitle

\section{Introduction}\label{1}
Black holes have fascinated physicists for a long time, not only because of their amazing properties in general relativity but also because of their connection to other fields of physics, such as quantum mechanics and thermodynamics \cite{Chandrasekhar:1931ih,Hawking:1975vcx,Bekenstein:1973ur}. Black hole thermodynamics is an intriguing area of research in theoretical physics. The idea that black holes have an entropy proportional to their horizon area was proposed by Bekenstein \cite{Bekenstein:1973ur} and verified by Hawking \cite{Hawking:1975vcx}, who showed that black holes emit thermal radiation with a temperature proportional to their surface gravity. This led to the definition of the Hawking temperature ($T$), the Bekenstein-Hawking entropy ($S_{BH}$), and the first law of black hole thermodynamics $dM=TdS_{BH}$, where $M$ is the mass of the black hole\cite{Bekenstein:2008smd}.

Since then, black hole thermodynamics has been developed significantly, and additional thermodynamic quantities have been introduced, such as the heat capacity, pressure, and chemical potential \cite{Brown:1994gs,Kubiznak:2016qmn}. Black hole thermodynamics is now studied not only in classical general relativity but also in various modified gravity theories, including 2-dimensional dilaton gravity, Lovelock gravity, deformed Horava-Lifshitz gravity and a class of black holes in de Rham, Gabadadze, and Tolley massive gravity\cite{Aman:2006mn,Grumiller:2007ju,grumiller2022generalized,Youm:1999xn,Myers:1988ze,Ghosh:2015cva,Wei:2010yw}.

Thermodynamic geometry is a mathematical framework that explores the geometry of the equilibrium states of a thermodynamic system using its thermodynamic variables. The stability of the system depends on the curvature of the geometry, which also reveals other thermodynamic properties. The curvature of the thermodynamic space is called thermodynamic curvature and it measures how the system's components interact statistically. For example, the thermodynamic space of an ideal classical gas is flat (zero curvature), while the thermodynamic space of an ideal quantum Bose (Fermi) gas is curved positively (negatively).The curvature as a function of thermodynamic variables has some singularities so that a kind of phase transition occurs at each critical point \cite{Ruppeiner:1995zz,janyszek1990riemannian,mirza2009nonperturbative,mirza2011condensation,mehri2020thermodynamic,ebadi2022thermodynamic}.

Thermodynamic geometry has been widely applied to black hole studies, especially for AdS geometries in the context of AdS/CFT correspondence \cite{Quevedo_2008,Larranaga2011wd,Quevedo2016,Wei:2015iwa}. The field of thermodynamic geometry has many applications in different physical scenarios of black holes. For example, it has been used to investigate the phase transitions and critical phenomena in black holes \cite{Wei:2012ui,Sahay:2010tx}, to understand the microscopic structure of black holes \cite{Wang:2019cax,Wei:2010yw}, to examine the critical behavior of black holes \cite{Shen:2005nu,myung2008ruppeiner,Rafiee:2021hyj}, and to explore the relation between the thermodynamic geometry and the holographic complexity\cite{Babaei-Aghbolagh:2022xcy}. The Ruppeiner metric, for instance, has been employed for various black hole models in different settings, such as AdS/CFT black holes \cite{Wei:2015iwa}, black holes in massive gravity \cite{Wu:2020fij}, and black holes in extended phase space \cite{Zhang:2015ova}, among others. Moreover, the instability of the metric indicates the occurrence of phase transitions, such as the Hawking-Page phase transition in anti-de Sitter space-time\cite{Hawking:1982dh}. Different geometric methods have been used to analyze the thermodynamic compatibility of different thermodynamic potentials, such as pressure and entropy, in black hole physics \cite{Dolan:2015xta,Kastor:2009wy,Niu:2011tb,GUO2022115839}.

JT gravity\cite{Teitelboim:1983ux,Jackiw:1984je} is a two-dimensional (2D) theory of dilaton gravity and the thermodynamics of JT black hole has been investigated widely \cite{Lemos1996ThermodynamicsOT,Youm:1999xn,Pedraza:2021cvx}. Recently, a deformed  version of JT gravity (dJT) has been proposed, which includes some generalized non-linear potentials of dilaton field\cite{Maldacena:2016hyu,Maxfield:2020ale,grumiller2022generalized,nejati2023jackiw}.
This model has attracted a lot of attention because of its implications for quantum gravity, holography, string theory, and condensed matter systems, especially for black hole physics. dJT gravity introduces a new coupling parameter that allows us to explore different physical scenarios by adding extra interactions between the dilaton field and matter. Previous studies have examined various features of dJT gravity, such as its quantization, classical solutions, thermodynamics, holographic dualities, and heavy-light spectra \cite{Mertens2018, Concha2020, Farquet2019, Cotler2017, Saad2019,Nojiri_2022}. This model is also interesting because of its relation to the Sachdev–Ye–Kitae (SYK) model in holographic AdS/CFT duality, which describes quantum matter out of equilibrium \cite{sachdev1993gapless,sachdev2010holographic,Kitaev2015,afshar2020flat,louw2023matching}.

We use thermodynamic geometry to study the properties of charged black holes in 2D dJT gravity, which is a modified version of JT gravity with extra couplings between the dilaton field and matter. This approach can help us understand the thermodynamic features of dJT gravity and its connections to other physical phenomena. Thermodynamic phase space of a black holes in dJT gravity has been considered and shown some promising results \cite{Cao:2021upq}.
The properties of black holes in dJT gravity have been the subject of several studies. For instance, in Ref.~\cite{Lu:2022tmt}, the behavior of the Page curve using the island paradigm was studied by considering the entanglement island in a dJT black hole in the presence of the phase transition. In Ref.~\cite{Johnson:2020lns} some puzzles about the phase transition and non perturbation effects were solved  and in Ref.~\cite{Alishahiha:2020jko},  the partition function and free energy of the model was obtained. Other studies have also investigated interesting phenomena such as the emergence of complexity in the model\cite{Alishahiha2019}. Also, the exact solution of dJT gravity for AdS and non-AdS cases has been inbestigated\cite{Momeni:2020zkx,momeni2021real}.

As main interests in this paper, we first study the phase structure of the charged black holes by finding the thermodynamic variables and their corresponding critical behaviors. The continuity of charge and entropy in one hand and the discontinuity in heat capacity for a critical value of entropy in the other hand rule out the existence of a first-order phase transition. However, we investigate the nature of the phase transition by examining the two Ehrenfest like equations for the charged black holes. This scheme in the case of Kerr black holes has been proposed in Ref.~\cite{banerjee2010glassy}. If a phase transition is of second-order, then the two Ehrenfest’s equations must be satisfied. Our numerical analysis for different values of constant parameters in dJT gravity reveals that the two Ehrenfest’s equation are satisfied and we have a second-order phase transition. Then, we will consider the thermodynamic geometry in three different approaches; Weinhold, Ruppeiner and new thermodynamic geometry. We show that the critical points of heat capacities are coincident with singularities of thermodynamic curvatures in the phase space.

The organization of the paper is as follows:  In section \ref{2} we review the structure of dJT gravity by introducing a coupling to the Maxwell action. We find the critical values of temperature, entropy and charge for the AdS black holes in dJT theory. Section \ref{3} delves deeply into the examination of heat capacities, expansion coefficient and isothermal compressibility at critical points. This section offers significant insights into the phase transition of the charged black hole solution in dJT gravity. This fact is considered by checking the Ehrenfest’s equations in the phase space.
Section \ref{4} presents an intriguing thermodynamic geometry approach that examines the thermodynamic curvature and divergences of heat capacities at constant charge and  chemical potential which are respectively denoted by $C_{Q}$ and $ C_{\Phi}$.  The approach involves various geometric methods and provides insightful comparisons.
Finally, we summarized the paper with some concluding remarks in Section \ref{5}.

\section{charged black hole in dJT gravity}\label{2}

In the following, we will consider a dJT gravity including a dilatonic potential denoted by function $V(\phi)$ so that in the asymptotic limit, the black hole solution is asymptotically $AdS_2$ as well as the black holes in JT gravity. Therefore, we use $V(\phi) = 2\phi + U(\phi)$, where $U(\phi)$ adds a perturbation to JT gravity. In these theories, as shown in \citep{Witten:2020ert}, there is a possibility of a first-order phase transition, like the Hawking-Page transition between two black holes - one with negative heat capacity and higher temperature, and the other with positive heat capacity and lower free energy.

The thermodynamics of the holographic dual to dJT gravity -- a complex SYK model-- has been investigated in Ref.~\cite{Cao:2021upq}. In this paper, we consider a 2D dJT gravity coupled to a Maxwell field $\mathcal{A_{\mu}}$ which is described by the following action
\begin{eqnarray}\label{action}
S_{\rm dJT}&=&\frac{\phi_0}{2}\bigg(\int_{\mathcal{M}}d^2 x \sqrt{-g}R+2\int_{\partial{M}}\sqrt{-h}K d \tau\bigg)\nonumber\\ &+&\frac{1}{2}\int_{\mathcal{M}}d^2 x \sqrt{-g}\bigg(\phi R+\frac{V(\phi)}{l^2}-\frac{1}{2}Z(\phi)\mathcal{F}^2\bigg)\nonumber\\ &+&\frac{1}{2}\int_{\partial{M}} d\tau \sqrt{-h} \,n_{\mu}Z(\phi)\mathcal{F}^{\mu\nu}\mathcal{A}_{\nu}+
\int_{\partial{\mathcal{M}}}\sqrt{-h} \phi K d\tau\nonumber\\ &+&\int_{\partial{\mathcal{M}}}\sqrt{-h} \mathcal{L}_{\rm ct}d\tau\,,
\end{eqnarray}
where K is the extrinsic curvature computed from the unit normal vector $n_{\mu}$ and  $\mathcal{F}_{\mu\nu}=\partial_{\mu} \mathcal{A}_{\nu}-\partial_{\nu} \mathcal{A}_{\mu}$ is the electromagnetic field strength and $Z(\phi)$ is a functional coupling of dilaton. $\mathcal{L}_{\rm ct}$ is a boundary counter term to have a finite on-shell action. The charged black hole solution for the equations of motion of the dJT gravity given by the action (\ref{action}) is specified by
\begin{eqnarray} \label{djt}
ds^2&=&-f(r)dt^2+\frac{dr^2}{f(r)}, \quad \phi(r)=r,\\
f(r)&=&\frac{r^2-r^2_{H}}{l^2}+\frac{a_0}{(1-\eta)l^2}\bigg(r^{1-\eta}-r_{H}^{1-\eta}\bigg)-\frac{Q^2}{(1-\xi)l^2}\bigg(r^{1-\xi}-r_{H}^{1-\xi}\bigg),  \nonumber\\
~Z(\phi)&=&r^{\xi},\quad \mathcal{A}_t(r)=  \Phi+\frac{Q}{1-\xi}r^{1-\xi}, \quad ~V(\phi(r))=2r+\frac{a_0}{r^{\eta}},\nonumber
\end{eqnarray}
where $\eta$, $\xi$, and $a_0$ are positive constants and $l$ is the radius of AdS space that for simplicity we set $l=1$. The chemical potential and entropy of the black hole on the event horizon $r_{H}$ are given by
\begin{eqnarray}\label{eq-S}
\Phi = \frac{Q}{\xi -1}r^{1-\xi}_H, \qquad S = 2\pi \phi_H=2 \pi r_ H,
\end{eqnarray}
whereas its ADM mass and temperature are as follows
\begin{eqnarray}\label{Tm}
M=\frac{1}{2} r_ H^2+\frac{a_0 }{2(1-\eta)} r_ H^{1-\eta}-\frac{Q^2 }{2(1-\xi)} r_ H^{1-\xi}, \,\,\,\,\,\,\,\,\,\,\,\, T= \frac{2 r^{\xi+1}_{H}+a_0 r^{\xi-\eta}_{H}-Q^2}{4 \pi r^{\xi}_{H}}.
\end{eqnarray}
More details about the calculation of the thermodynamic variables such as entropy, temperature and mass are given in supplemental materials of Ref.~\cite{Cao:2021upq}.
\subsection{Extremal Limit}
In general, the extremal limit is obtained when the temperature of the black hole goes to zero. In other words, zero temperature black holes are black holes that have zero Hawking temperature, which means they do not emit any thermal radiation. However, they may still have nonzero entropy and mass.
Using Eqs.~(\ref{eq-S}) and (\ref{Tm}), we obtain the temperature and the mass of charged black hole as explicit functions of the charge and entropy, i.e.,
\begin{eqnarray}\label{TM}
T=\frac{S}{4 \pi^2} + \frac{2^{\eta-2} a_0 \pi^{\eta-1}}{ S^{\eta}} - \frac{ 2^{\xi-2} \pi^{ \xi-1} Q^2}{ S^{\xi}}\,,\qquad
M=\frac{S^2}{8 \pi^2} -  \frac{2^{ \eta-2} a_0 \pi^{ \eta-1}}{(\eta-1) S^{\eta-1}} + \frac{2^{ \xi-2} \pi^{ \xi-1} Q^2}{( \xi-1) S^{\xi-1}}\,.
\end{eqnarray}
It is shown that in the extremal limit, we have a charge given by the following relation
\begin{eqnarray}\label{QE}
Q^2=Q^2_{E}=\frac{S^{\xi} (2^{\eta} a_0 \pi^{1 + \eta} + S^{1 + \eta})}{2^{\xi} \pi^{1 + \xi} S^{\eta}},
\end{eqnarray}
and a mass corresponding to this charge as
\begin{equation}\label{eq-ME}
M_{E}= \frac{2^{ \eta -2} a_0 \pi^{ \eta -1} (\eta -  \xi)}{S^{\eta -1} ( \eta -1) ( \xi -1)} + \frac{S^2 (1 + \xi)}{8 \pi^2 ( \xi -1)}.
\end{equation}
In fact this is a minimum possible mass that a black hole can have for a given charge $Q_{E}$. According to Eqs.~(\ref{TM}) and (\ref{eq-ME}), one can recast the temperature  in the form
\begin{equation}\label{eq-TME}
T= \frac{ (1 - \xi)}{S}\,( M-M_{E}) \,,
\end{equation}
which is manifestly disappeared in the extremal limit. This temperature also shows that zero temperature black holes have non-vanishing entropy which might be due to the quantum effects or higher curvature corrections.
\subsection{Critical Temperature}\label{22}
In black hole thermodynamics, the critical points are the locations at which a phase transition occurs. This effect in general happens for the case of charged black holes which are identified with a liquid–gas system in the extended phase space, for instance a van
der Waals gas proposed in Ref.~\cite{Chamblin:1999tk}. The van der Waals equation of state, which describes the behavior of real gases, shows features that resemble black hole thermodynamics, making it a convenient tool for studying the heat capacity of black holes.
 The critical behavior of AdS black holes in an extended phase space, including pressure and volume as thermodynamic variables was proposed in Ref.~\cite{dolan2011pressure}. This $P-V$ criticality for both the van der Waals gas and the spherically RN-AdS black holes was investigated in Ref.~\cite{Kubiznak:2012wp}. In fact, in the $P-V$ diagram the critical point occurs when $P$ has an inflection point, i.e. $\partial P/\partial V=0$ and $\partial^2 P/\partial V^2=0$. However, it has also been investigated that this phase transition can be observed in other inflection ponits of an extended phase space. For example in Ref.~\cite{Rafiee:2021hyj}, the behavior of charged AdS black hole solution in Einstein-Maxwell-AdS gravity has been studied in the $T-S$ phase diagram for an inflection point  $\partial T/\partial S=0$ and $\partial^2 T/\partial S^2=0$. Practically, the former criticality is related to some isothermal behavior while in the latter case the isocharge curves are studied.

On the other hand, in the thermodynamic phase space, the critical temperature denotes the threshold temperature at which a phase transition occurs. This transition is commonly observed in the context of charged black holes, where a transition from a stable, large black hole to an unstable, small black hole occurs above a critical temperature. The critical temperature marks the point where the thermodynamic properties of the black hole undergo a significant change. In the following, we will explore the $T-S$ criticality of 2D charged black holes in dJT gravity described by (\ref{djt}) at an inflection point where
\begin{eqnarray} \label{inf}
 \left(\frac{\partial T}{\partial S}\right)_{Q}=0,\qquad \left(\frac{\partial^2 T}{\partial S^2}\right)_{Q}=0.
\end{eqnarray}
 We will also investigate the behavior of the heat capacities near the critical points in the next section. Solving the above equations for the temperature in (\ref{TM}), we find the following critical values of the charge and entropy of black hole
\begin{eqnarray} \label{qsc}
S_c=\bigl( \frac{2^{\eta} a_0 \pi^{1 + \eta} \eta (  \xi-\eta)}{1 + \xi}\bigr)^{\frac{1}{1 + \eta}},\qquad Q^2_c=\frac{2^{\frac{\eta -  \xi}{1 + \eta}} a_0^{\frac{1 + \xi}{1 + \eta}} \eta^{\frac{1 + \xi}{1 + \eta}} (1 + \eta) ( \xi- \eta )^{- \frac{\eta -  \xi}{1 + \eta}}}{\xi (1 + \xi)^{\frac{1 + \xi}{1 + \eta}}}.
\end{eqnarray}
Then, substituting these values in the temperature (\ref{TM}) we obtain the critical temperature as follows
\begin{eqnarray} \label{tc}
T_c=\frac{a_0^{\frac{1}{1 + \eta}} (1 + \eta) (1 + \xi)^{\frac{\eta}{1 + \eta}} (\xi- \eta )^{\frac{1}{1 + \eta}}}{2^{\frac{2 + \eta}{1 + \eta}} \pi \eta^{\frac{\eta}{1 + \eta}} \xi}.
\end{eqnarray}
As is obvious all the critical values depend on the constant parameters of the theory such that when $\xi=\eta$ all of them vanish and there is no phase transition. In order to have physical quantities $\xi$ should be also greater than $\eta$. In fact, the qualitative features of the phase structure depend on these parameters and we will
investigate them in different situations.

\begin{figure}[h]
	\begin{subfigure}{0.45\textwidth}\includegraphics[width=\textwidth]{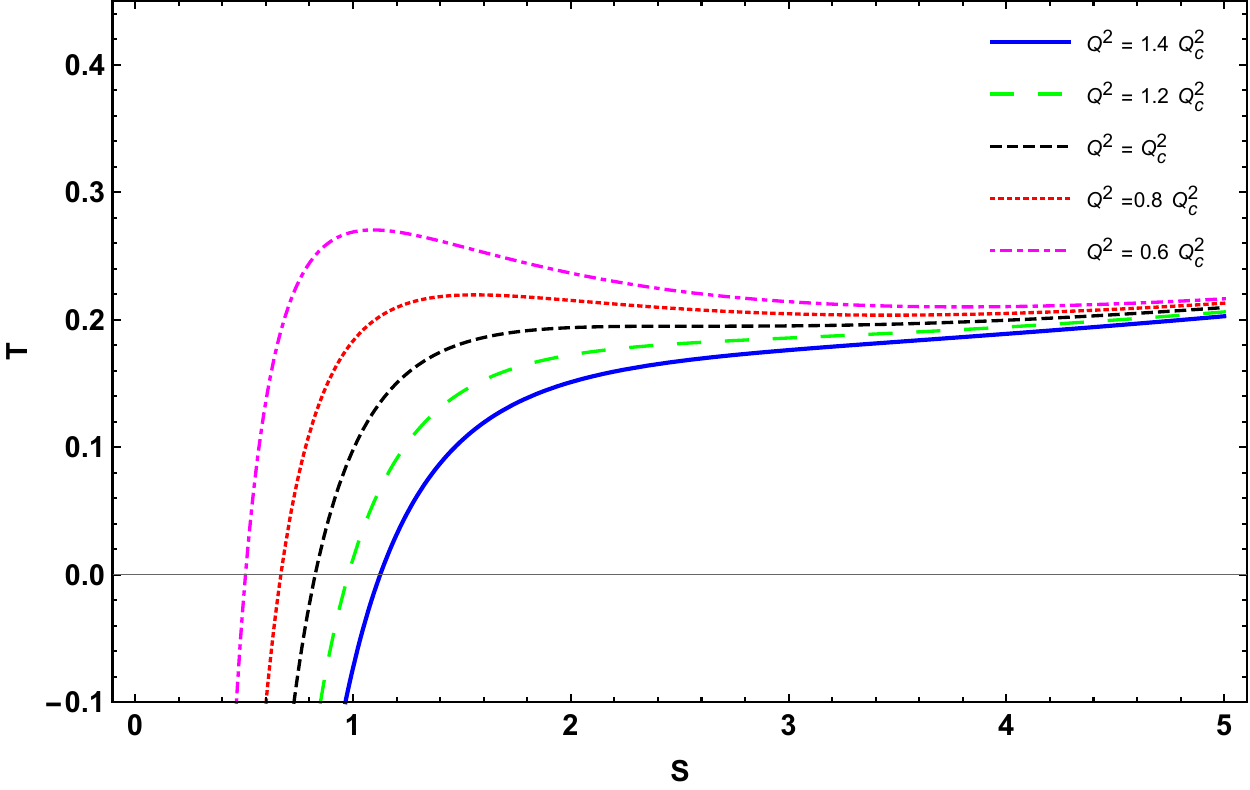}
		\caption{}
		\label{fig:1-1}
	\end{subfigure}
	\begin{subfigure}{0.45\textwidth}\includegraphics[width=\textwidth]{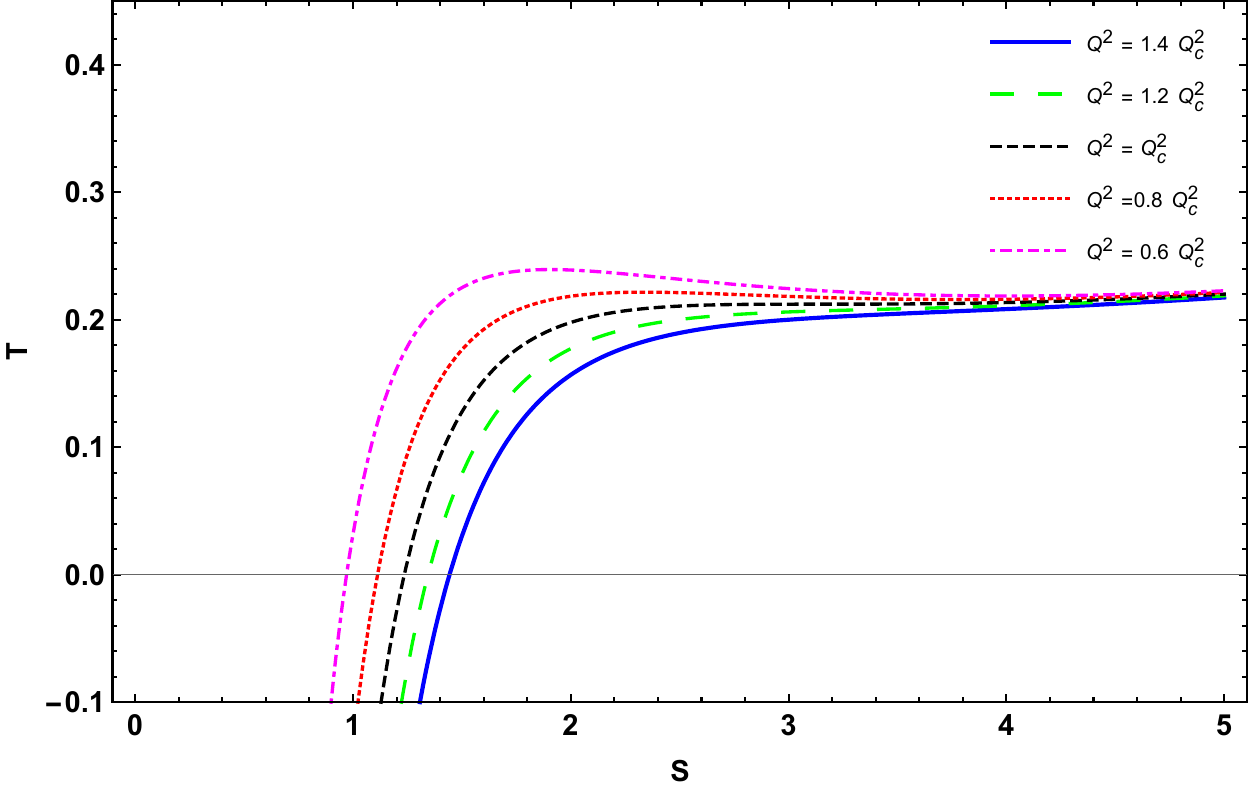}
		\caption{}
		\label{fig:1-2}
	\end{subfigure}	\begin{subfigure}{0.45\textwidth}\includegraphics[width=\textwidth]{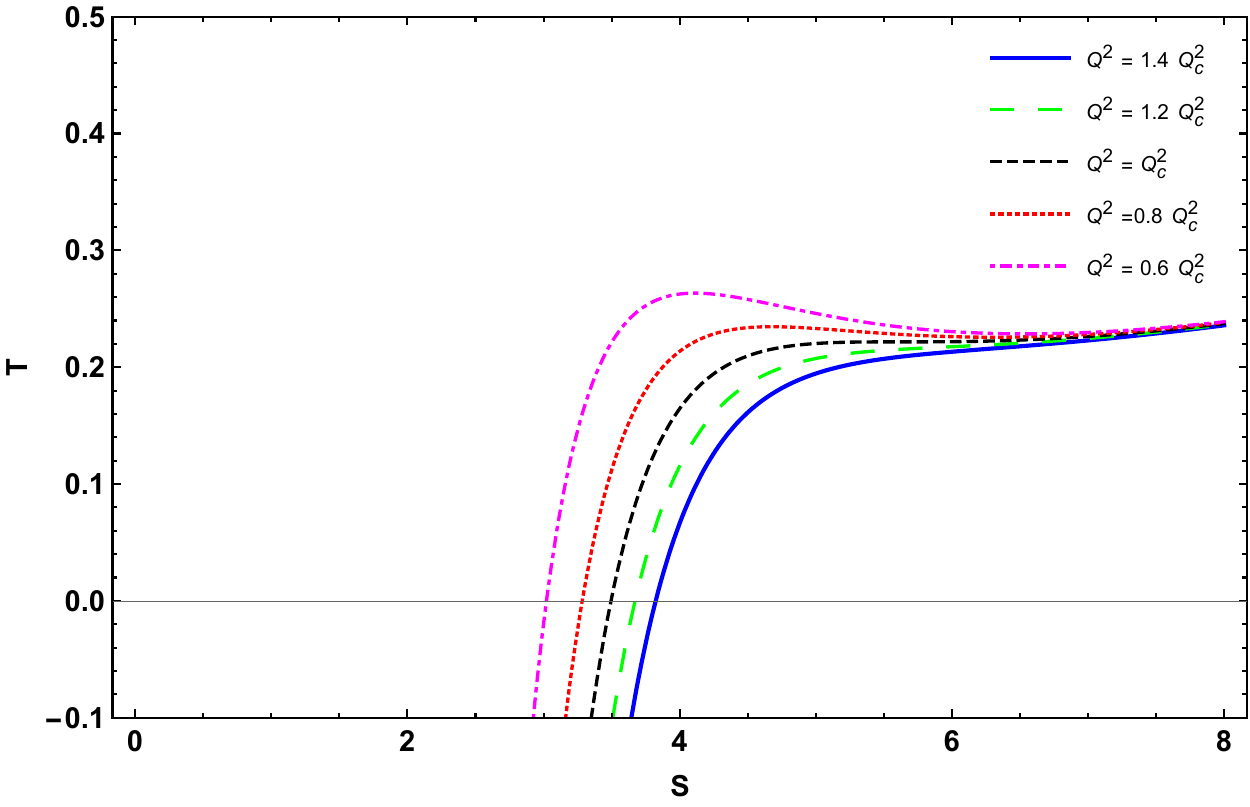}
	\caption{}
	\label{fig:1-3}
\end{subfigure}	\begin{subfigure}{0.45\textwidth}\includegraphics[width=\textwidth]{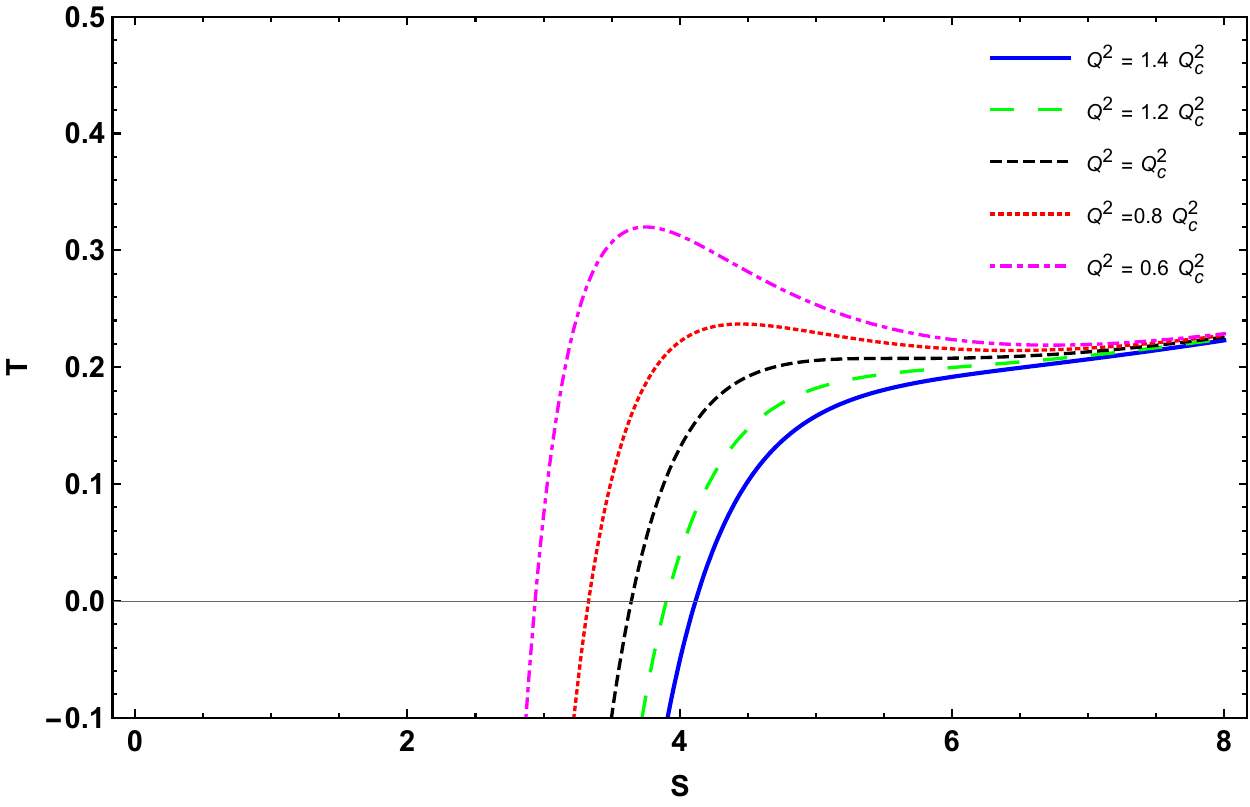}
\caption{}
\label{fig:1-4}
\end{subfigure}
	\caption{The isocharge $(T,S)$ curves of charged AdS black holes in dJT gravity for $a_0=1$; (\ref{fig:1-1}) $\eta=1$ and $\xi=2$, (\ref{fig:1-2}) $\eta=1$ and $\xi=3$, (\ref{fig:1-3}) $\eta=3$ and $\xi=6$, (\ref{fig:1-4}) $\eta=4$ and $\xi=6$.}
	\label{fig:1}
\end{figure}

In Fig.~\ref{fig:1}, we have plotted the temperature as a function of entropy for different values of charge. The value of charge for the isocharge curves decreases from top to bottom. The points that curves cross the horizontal axis correspond to extremal limit. In accordance with Eq.~(\ref{eq-TME}), these zero temperatures have explicitly non-zero entropies. One can also observe that the temperature has a local maximum when $Q>Q_c$ for all the values of $\eta$ and $\xi$ so that for fixed $\eta$ when we increase the value of $\xi$ the local peak decreases, in contrast for fixed $\xi$, by increasing $\eta$ the value of maximum increases. The first case is related to increasing of the strength of coupling dilaton to Maxwell theory, i.e. $Z(\phi)$, while in the latter the deformation function $U(\phi)$ is suppressed. The other important point that can be seen from these figures is that for a fixed value of charge $Q$, there is no first-order phase transition since the curves have no discontinuity for different values of entropy.

\begin{figure}[h]
	\begin{subfigure}{0.45\textwidth}\includegraphics[width=\textwidth]{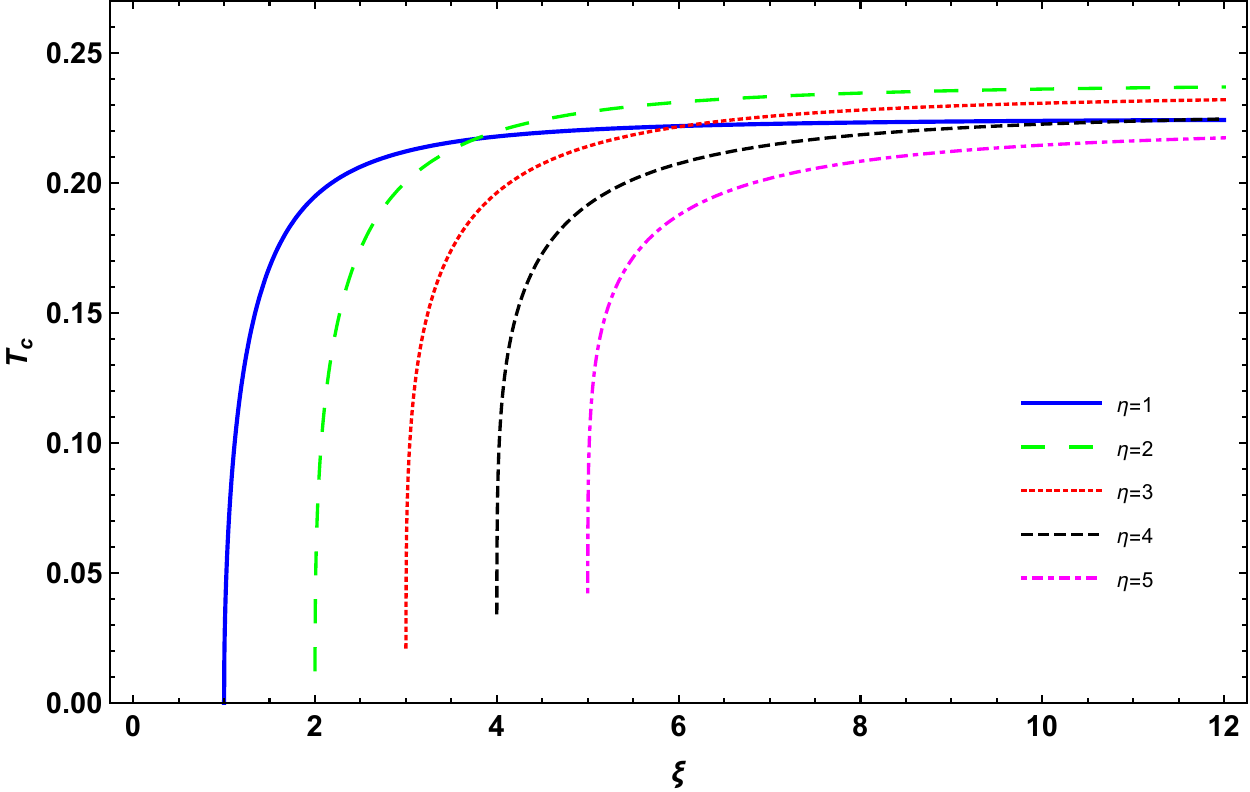}
		\caption{}
		\label{fig:TSQ1-1}
	\end{subfigure}
 \begin{subfigure}{0.45\textwidth}\includegraphics[width=\textwidth]{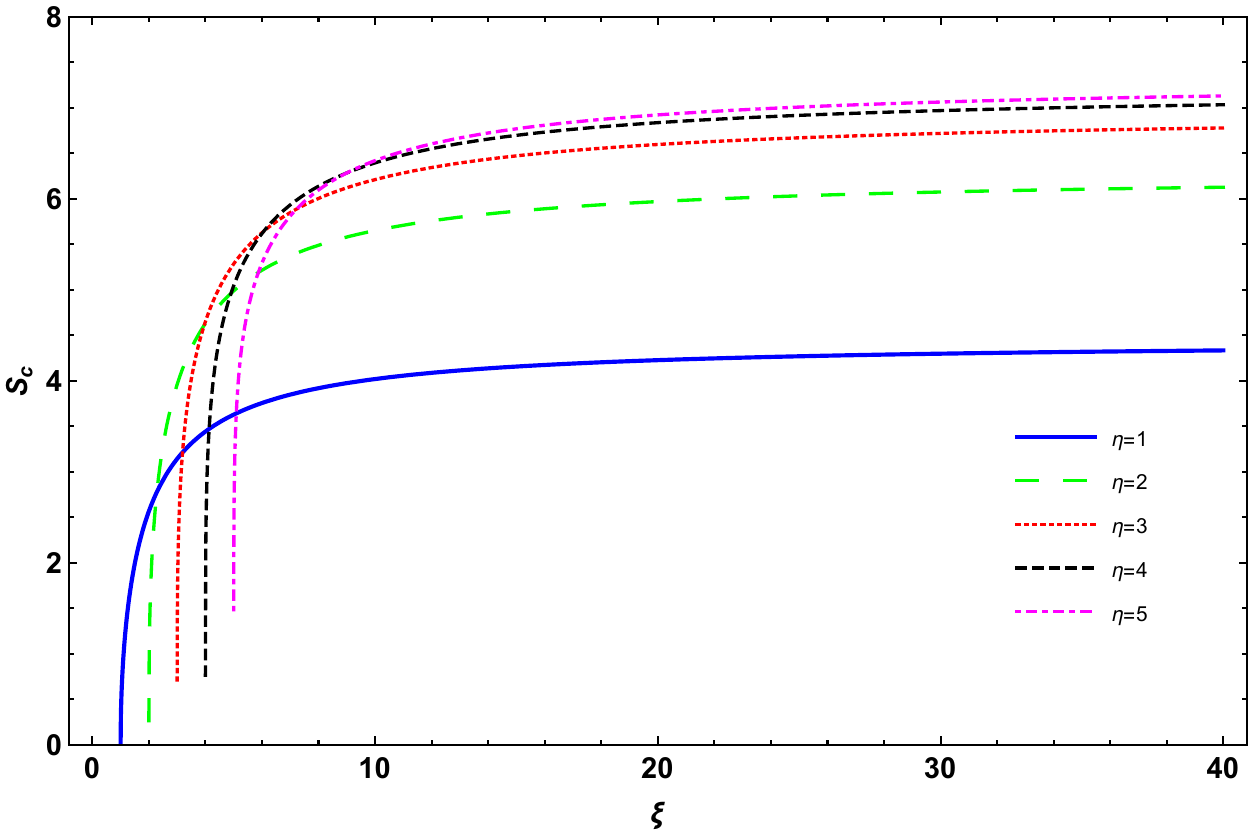}
		\caption{}
		\label{fig:TSQ1-2}
	\end{subfigure}
	\begin{subfigure}{0.45\textwidth}\includegraphics[width=\textwidth]{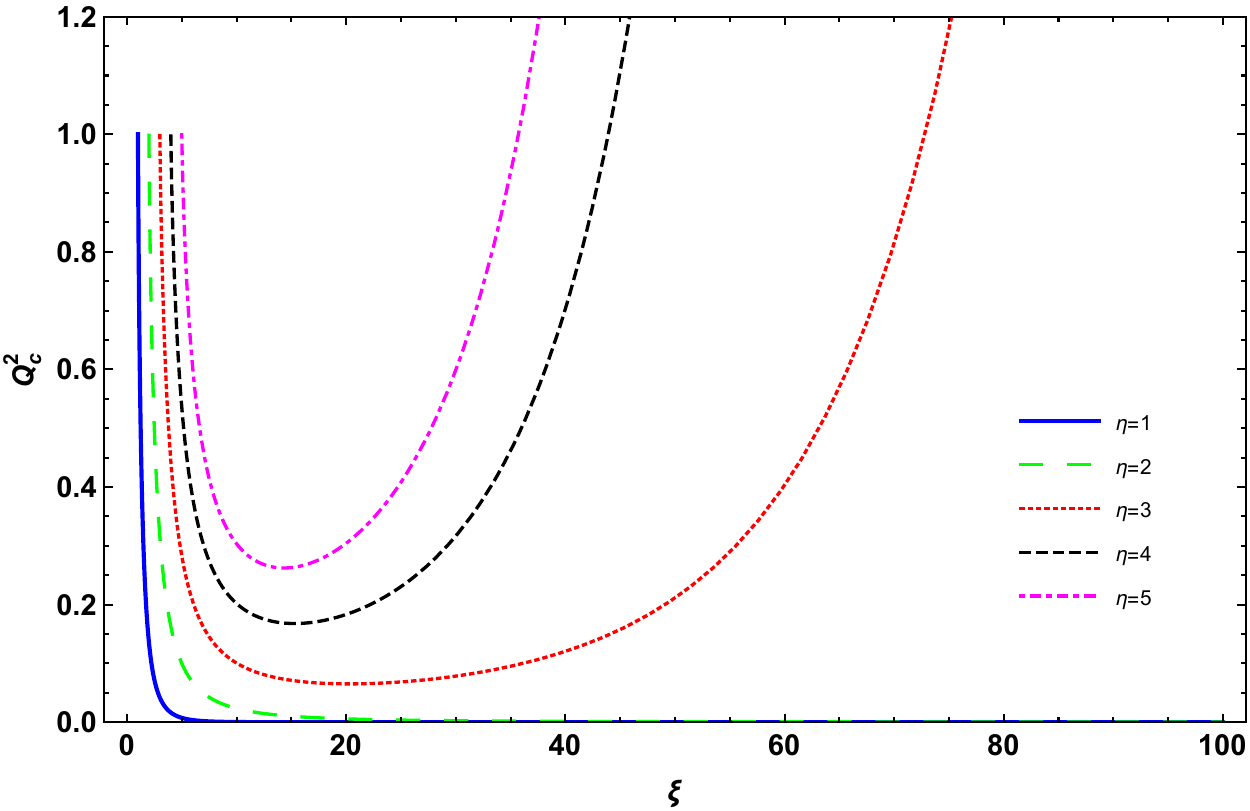}
		\caption{}
		\label{fig:TSQ1-3}
	\end{subfigure}
	\caption{ The behavior of $T_c$, $S_c$, and $Q^2_c$ vs. $\xi$ for different values of $\eta$ when $a_0=1$.}
	\label{fig:1TSQ}
\end{figure}

One can also study the behavior of the critical quantities given in Eqs.~(\ref{qsc}) and (\ref{tc}) in terms of different values of parameters by the constraint $\xi>\eta>0$. For instance, we list the critical temperature $T_c$ and critical entropy $S_c$ via different values of $\eta$ in Tab. (\ref{tab1}). Here, the increase of $\eta$ corresponds to quenching of the deformation potential $V(\phi)$. Therefore, to better understand how they change under this quench we have plotted them in Fig.~\ref{fig:1TSQ}.
\begin{table}[h!]
	\begin{center}
		\renewcommand*{\arraystretch}{1.5}
		\begin{tabular}{|c|c|c|}\hline
			$\eta$ & $T_c$  &$S_c$   \\ \hline
			$\eta=1$ &  $\frac{\sqrt{\xi-1} \sqrt{ \xi+1}}{\sqrt{2} \pi \xi}$ & $\pi \sqrt{2}\sqrt{\frac{  \xi-1}{ \xi+1}} $ \\  \hline
			$\eta=2$ & $\frac{3 ( \xi-2)^{1/3} ( \xi+1)^{2/3}}{4 \pi \xi}$  &$ 2 \pi ( \frac{ \xi-2}{ \xi+1})^{1/3}$ \\ \hline
			$\eta=3$ & $\frac{2^{3/4} (\xi-3)^{1/4} (\xi+1)^{3/4}}{3^{3/4} \pi \xi}$ &$2^{3/4} 3^{1/4} \pi (\frac{  \xi-3}{ \xi+1})^{1/4}$   \\ \hline
			$\eta=4$ & $\frac{5 ( \xi-4)^{1/5} ( \xi+1)^{4/5}}{4^{4/5} 2^{6/5} \pi \xi}$  & $2\times 2^{1/5} \pi ( \frac{  \xi-4}{\xi+1})^{1/5}$  \\ \hline
			$\eta=5$& $ \frac{3 (\xi-5)^{1/6} (\xi+1)^{5/6}}{2^{1/6} 5^{5/6} \pi \xi}$ &$2^{5/6} 5^{1/6} \pi (\frac{ \xi-5}{ \xi+1})^{1/6}$  \\ \hline
		\end{tabular}
		\caption{ The critical values of temperature and entropy for different values of $\eta$ as functions of $\xi$ when $a_0=1$.  \label{tab1}}
	\end{center}
\end{table}

According to Figs.~\ref{fig:TSQ1-1} and \ref{fig:TSQ1-2}, both $T_c$ and $S_c$ grow fast and then approach to some asymptotic values for large $\xi$. In the following, we derive analytical expressions for the asymptotic values from Eqs.~(\ref{qsc}) and (\ref{tc}), i.e.,
\begin{eqnarray}
T_{c}\Bigg|_{\xi \to \infty}=\frac{\eta^{\frac{1}{1 + \eta}} (1 + \eta)}{2^{\frac{2 + \eta}{1 + \eta}} \pi \eta} \,,\qquad S_{c}\Bigg|_{\xi \to \infty}=2^{\frac{\eta}{1 + \eta}} \pi \eta^{\frac{1}{1 + \eta}}\,.
\end{eqnarray}
However, when both of the parameters $\xi \to \infty$ and $\eta \to \infty$, we obtain $T_{c}=\frac{1}{2 \pi }$ and $S_{c}=2 \pi$, which are consistent with the results computed in Ref.~\cite{cadoni1995nonsingular} for a special value of scalar field at the critical point.
Also all the plots in Fig.~\ref{fig:1TSQ} approve the condition that $\xi$ should be always greater than $\eta$.

On the other hand, it is shown in Fig.~\ref{fig:TSQ1-3} that the critical charge exhibits different behaviors depending on the values of the parameter $\eta$. When $\eta \leq 2$, by increasing $\xi$ the critical charge decreases as far as it approaches zero for larger value of $\xi$, while for $\eta > 2$, it first decreases and then increases as $\xi$ grows. Finally, it diverges in the large limit of $\xi$, i.e.,
\begin{eqnarray}
Q^2_{c}\Bigg|_{\xi \to \infty,\, \eta \le 2}=0 \,,\qquad Q^2_{c}\Bigg|_{\xi \to \infty,\, \eta >2}=\infty\,.
\end{eqnarray}

\section{phase transition}\label{3}
Black hole thermodynamics, which involves the analysis of heat capacity, has progressed with the development of various black hole models. A remarkable analogy that has been widely explored is the similarity between black holes and the van der Waals gas, as alluded in the previous section. In fact, this resemblance, which implies that the pressure-volume phase transition in the van der Waals gas is comparable to the black hole phase transition from a small to a large black hole, makes it a convenient tool for studying the heat capacity of black holes. In other words, the heat capacity of the black hole can be linked to the heat capacity of the van der Waals fluid, revealing important aspects of black hole thermodynamics. This correspondence has also led to the discovery of various fascinating phenomena, such as triple points, evaporation/thermodynamics, critical exponents and reentrant phase transitions, which have been investigated in recent papers \cite{grumiller2006ramifications,Wei:2019uqg,NaveenaKumara:2020biu,johnson2023specific, frassino2023reentrant}. On the other hand, the heat capacities of a black hole at some constant quantities are determining characteristics in studying the stability and phase transition in the black hole thermodynamics. In the following, we first obtain the heat capacities at constant electric charge and chemical potential for black holes in dJT gravity and investigate their critical behaviors, then show that these black holes undergo second-order phase transition by considering two Ehrenfest like equations.
\subsection{Heat Capacities}
The heat capacities at constant electric charge ($C_Q$) and chemical potential ($C_\Phi$), which are familiar analogues of the heat capacities at constant volume ($C_V$) and pressure ($C_P$) in fluid systems, determine the thermal stability of the
black holes. In fact, stability followed from non-negative heat capacities shows that black holes radiate at higher temperatures when they are smaller. The stability conditions in the case of charged AdS black holes have been initially considered in Ref.~\cite{chamblin1999holography}.
As a main interest in this paper, we investigate the heat capacity at constant charge in the phase space and its divergence at the critical point, which indicates that the system undergoes a phase transition at this point. This section aims to analyze the phase transition that happens in the charged black hole solution in the context of dJT gravity. This solution has a thermodynamic phase structure similar to the van der Waals-Maxwell, with a critical point where the heat capacity at constant charge diverges.
The divergence of the heat capacity at constant charge is usually associated with a phase transition in black hole thermodynamics. Moreover, it has been proposed in Ref.~\cite{Banerjee:2010bx} that there is another kind of phase transition for black holes that involves a divergence in the heat capacity at constant chemical potential.

\begin{figure}[h]
	\begin{subfigure}{0.45\textwidth}\includegraphics[width=\textwidth]{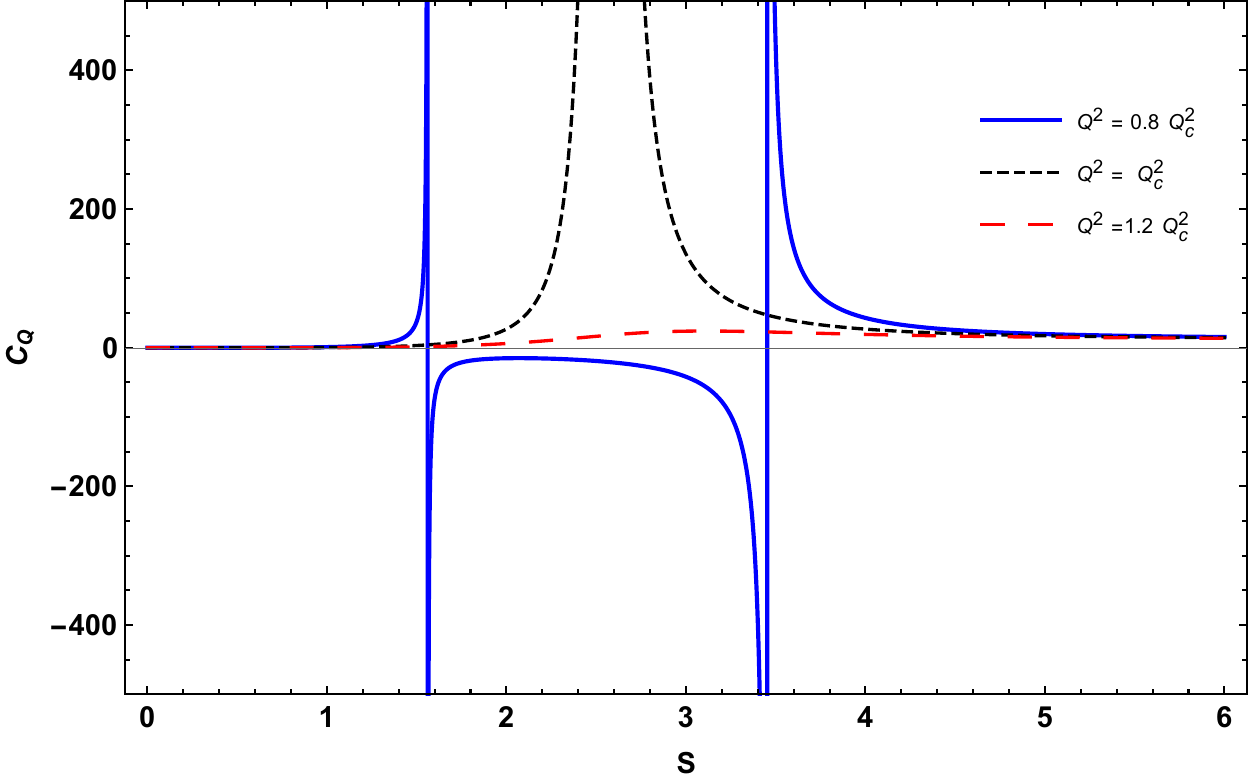}
		\caption{}
		\label{fig:2-1}
	\end{subfigure}
	\begin{subfigure}{0.45\textwidth}\includegraphics[width=\textwidth]{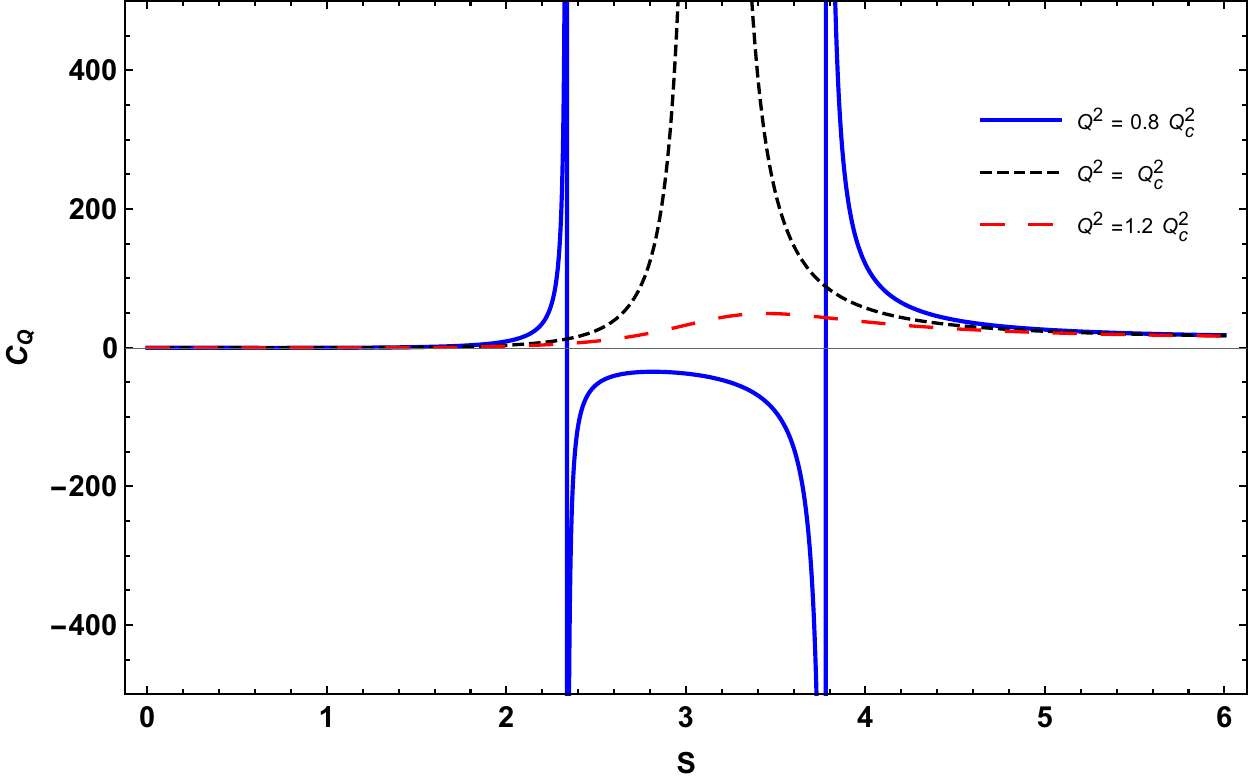}
		\caption{}
		\label{fig:2-2}
	\end{subfigure}
	\caption{Heat capacity at constant charge vs. entropy for different values of charge when $a_0=1$ and  $\eta=1\,;$    (\ref{fig:2-1})  $\xi=2$  and (\ref{fig:2-2}) $\xi=3$.}
	\label{fig:2}
\end{figure}

We employ the following Poisson's brackets and partial derivatives to find the heat capacity \cite{mansoori2015hessian}:
\begin{eqnarray}
{{\left( \frac{\partial f}{\partial g} \right)}_{h}}=\frac{{{\left\{ f,h \right\}}_{a,b}}}{{{\left\{ g,h \right\}}_{a,b}}}\,,
\end{eqnarray}
and
\begin{eqnarray}\label{eq16}
{{\{f,h\}}_{a,b}}={{\left( \frac{\partial f}{\partial a} \right)}_{b}}{{\left( \frac{\partial h}{\partial b} \right)}_{a}}-{{\left( \frac{\partial f}{\partial b} \right)}_{a}}{{\left( \frac{\partial h}{\partial a} \right)}_{b}}\,,
\end{eqnarray}
where $f$, $g$, and $h$ are functions of two variables $a$ and $b$. Then, the heat capacity at constant charge can be defined as follows
\begin{eqnarray}\label{eq17}
{{C}_{Q}}=T{{\left( \frac{\partial S}{\partial T} \right)}_{Q}}=\frac{T{{\left\{ S,Q \right\}}_{S,Q}}}{{{\left\{ T,Q \right\}}_{S,Q}}}\,.
\end{eqnarray}
Using the temperature in Eq.~(\ref{TM}) and substituting it into Eq.~(\ref{eq17}), we compute the heat capacity at constant charge explicitly given by
\begin{eqnarray}\label{eq18}
C_{Q}=\frac{S (2^{\eta} a_0 \pi^{1 + \eta} S^{\xi} + S^{1 + \eta + \xi} -  2^{\xi} \pi^{1 + \xi} S^{\eta} Q^2)}{S^{1 + \eta + \xi} -  2^{\eta} a_0 \pi^{1 + \eta} S^{\xi} \eta + 2^{\xi} \pi^{1 + \xi} S^{\eta} \xi Q^2}\,.
\end{eqnarray}
We have a keen interest in examining the divergence of the heat capacity at constant charge. As shown in Fig.~\ref{fig:2}, the heat capacity displays different behavior in three different regimes. It  has two divergent points for $Q^2<Q_{c}^2$. For $Q^2=Q_{c}^2$ one of the divergent points will be remained. For the case $Q^2>Q_{c}^2$, the heat capacity is an analytic function with respect to the entropy and there is no divergent point.

\begin{figure}[h!]
	\begin{subfigure}{0.45\textwidth}\includegraphics[width=\textwidth]{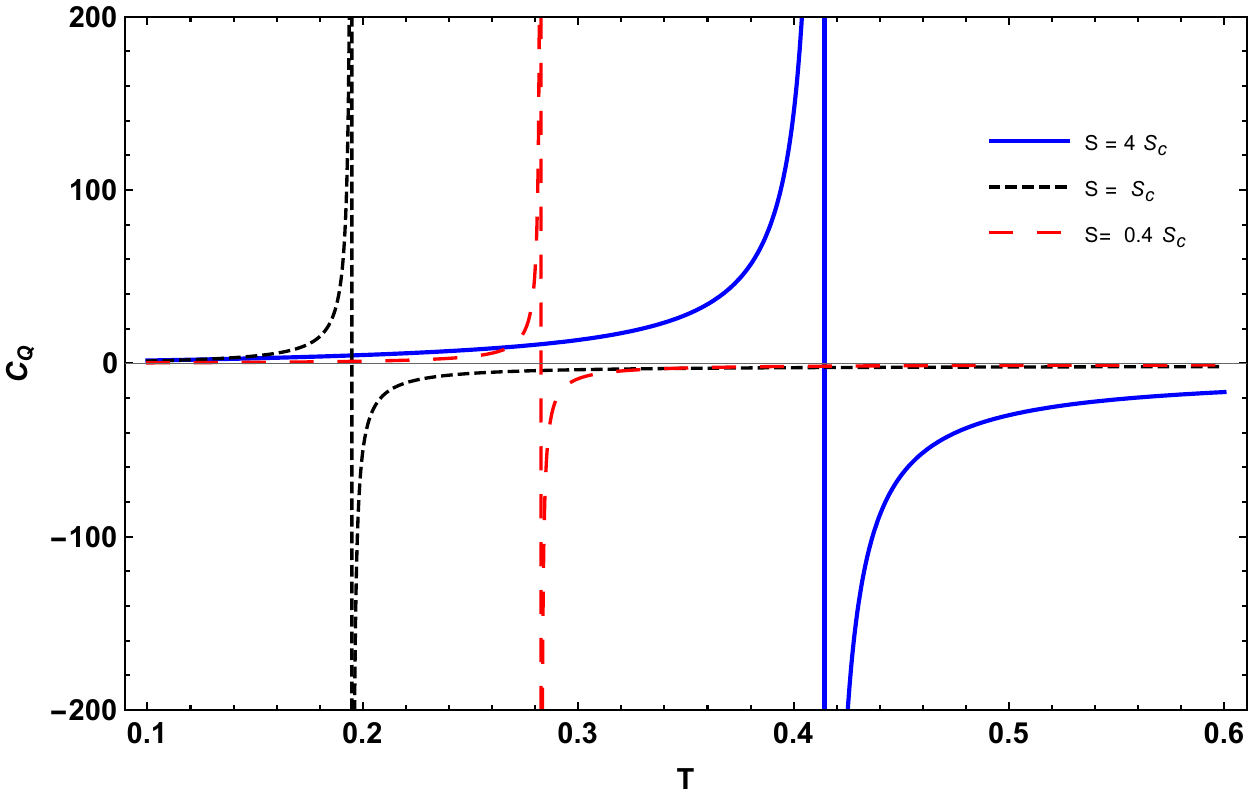}
		\caption{}
		\label{fig:3-1}
	\end{subfigure}
	\begin{subfigure}{0.45\textwidth}\includegraphics[width=\textwidth]{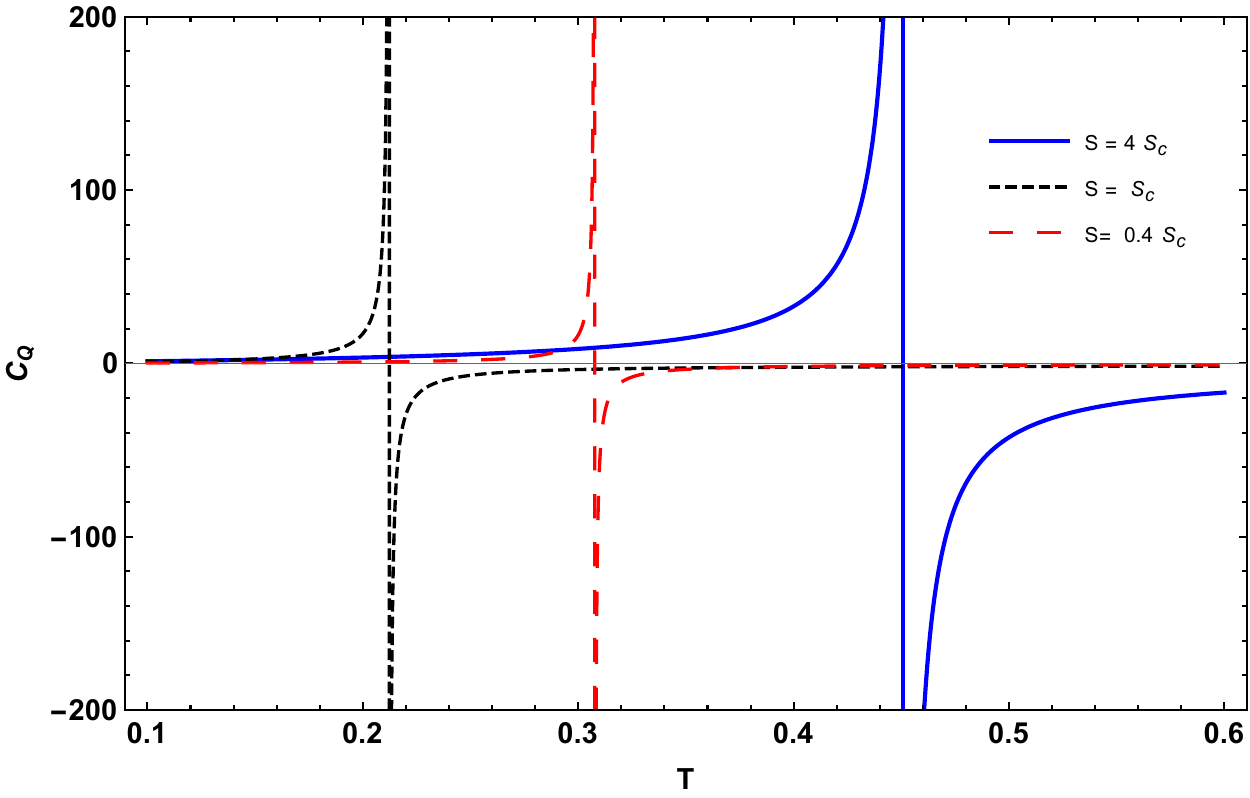}
		\caption{}
		\label{fig:3-2}
	\end{subfigure}
	\caption{Heat capacity at constant charge vs. temperature for different values of entropy when $a_0=1$ and  $\eta=1\,;$    (\ref{fig:3-1})  $\xi=2$  and (\ref{fig:3-2}) $\xi=3$.}
	\label{fig:3}
\end{figure}

We are also interested in finding how the heat capacity at constant charge varies with temperature. In this respect, using temperature in Eq.~(\ref{TM}) one can define $Q^2$ as a function of $T$ and $S$ by the following relation
\begin{equation}\label{eq-TM}
Q^2= \frac{S^{\xi} ( 2^{\eta} a_0 \pi^{1 + \eta} +  S^{1 + \eta} - 4 \pi^2 S^{\eta} T)}{2^{\xi} \pi^{1 + \xi} S^{\eta}}\,.
\end{equation}
Then, substituting Eq.~(\ref{eq-TM}) into Eq.~(\ref{eq18}) we can recast the heat capacity at constant charge in terms of temperature $T_{c}(S)=\frac{S^{1 + \eta} (1 + \xi)+2^{\eta} a_0 \pi^{1 + \eta} ( \xi-\eta  ) }{4 \pi^2 S^{\eta} \xi}$  as follows
\begin{eqnarray} \label{cq}
C_{Q}=-\frac{S\,\, T}{\xi (T  - T_{c}(S) )}\,.
\end{eqnarray}
The  temperature $T_{c}(S)$ corresponds to the point where the  derivative of temperature with respect to entropy vanishes. This implies a singular behavior of the heat capacity at constant charge as $T_{c}(S)$ approaches the critical temperature $T_c$. Indeed, the heat capacity follows a power-law behavior of the form $C_{Q}\sim {(T  - T_{c} )} ^{-\alpha}$ with $\alpha=1$. We have plotted $C_{Q}$ vs. $T$ for different values of entropy in Fig.~\ref{fig:3}. It is observed that the heat capacity has a critical behavior for each value of the entropy. In other words, $C_{Q}$  becomes infinite for each value of the entropy at some critical temperature $T_{c}$.

The heat capacity at constant chemical potential is also a measure of the sensitivity of the black hole's energy to changes in the temperature. The investigation of this quantity has important implications for understanding the stability and phase transitions of the charged black holes. For example, in their study of black holes in AdS spacetime, Hawking and Page \cite{Hawking:1982dh} showed that such black holes have positive heat capacity and can be in stable equilibrium with thermal radiation at a fixed temperature. Similarly, in Ref.~\cite{Chamblin:1999tk}, the heat capacities at constant charge and chemical potential of charged AdS black holes have been studied and multiple divergences in the heat capacities have been found that indicating phase transitions. The thermodynamics of black holes was further extended  by introducing cosmological pressure and proposed a modified definition for the heat capacity at constant chemical potential\cite{Kubiznak:2014zwa}. In the rest of this section we will investigate the heat capacity at constant potential for charged black holes of dJT gravity.

Heat capacity at constant chemical potential is defined  by
\begin{eqnarray}\label{cphi}
{{C}_{  \Phi}}=T{{\left( \frac{\partial S}{\partial T} \right)}_{  \Phi}}=\frac{T{{\left\{ S,  \Phi \right\}}_{S,Q}}}{{{\left\{ T,  \Phi \right\}}_{S,Q}}}\,.
\end{eqnarray}
Using the temperature (\ref{TM}) and chemical potential (\ref{eq-S}) in the above definition we obtain
\begin{eqnarray} \label{cphi}
C_{  \Phi}=\frac{(2 \pi)^{\xi} S (2^{\eta} a_0 \pi^{1 + \eta} S^{\xi} + S^{1 + \eta + \xi} -  2^{\xi} \pi^{1 + \xi} S^{\eta} Q^2)}{(2 \pi)^{\xi} S^{1 + \eta + \xi} -  2^{\eta + \xi} a_0 \pi^{1 + \eta + \xi} S^{\xi} \eta -  2^{2 \xi} \pi^{1 + 2 \xi} S^{\eta} ( \xi -2) Q^2}\,.
\end{eqnarray}
Similar to the heat capacity at constant charge in Eq.~(\ref{cq}), we can define the heat capacity at constant potential given by Eq.~(\ref{cphi}) in terms of a critical temperature, that is
\begin{eqnarray}
C_{  \Phi}=\frac{S \,\,T}{ ( \xi-2)\,\,(T - T'_{c}(S))}\,,
\end{eqnarray}
which shows a divergence at some critical temperature given by
\begin{eqnarray} \label{Tcp}
T'_{c}(S)=\frac{S^{1 + \eta} ( \xi-3) + a_0 \pi \bigl( (2 \pi)^{\eta} (\eta + \xi)- 2^{1 + \eta} \pi^{\eta} \bigr)}{4 \pi^2 \,\,( \xi-2)\,\,S^{\eta} }\,.
\end{eqnarray}

\begin{figure}
	\centerline{\includegraphics[scale=.40]{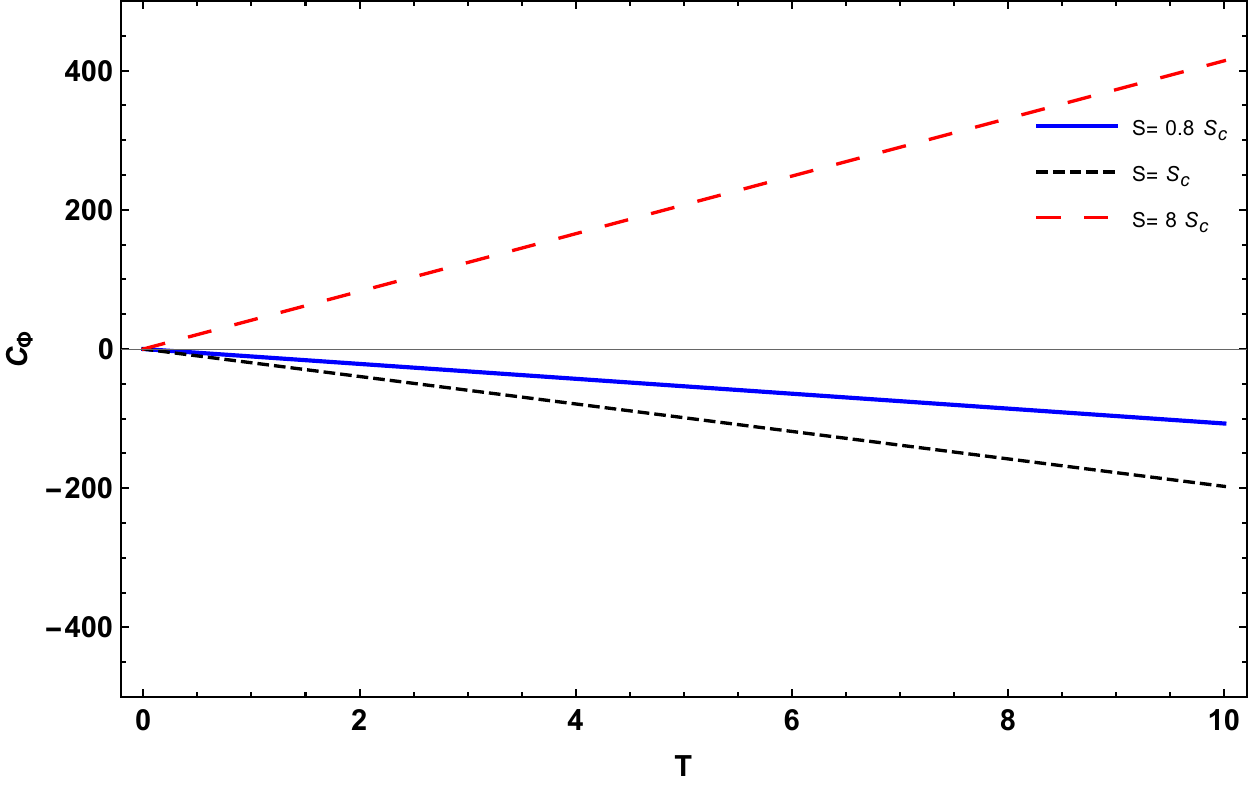}}
	\caption{Heat capacity at constant chemical potential vs. temperature for different values of entropy when $a_0=1$ and  $\xi=2$.}\label{fig:40}
\end{figure}

We want to explore the linear behavior of the heat capacity at constant chemical potential when $\xi=2$. In this case, the heat capacity can be expressed as a function of temperature as follows
\begin{eqnarray} \label{Tcp2}
C_{  \Phi}=\frac{4 \pi^2 }{1 -  \frac{2^{\eta} a_0 \pi^{1 + \eta} \eta}{S^{1 + \eta}}}T\,,
\end{eqnarray}
so that at the critical value of entropy given by Eq. (\ref{qsc}) it reduces to $C_{  \Phi}=\frac{4 \pi^2 (\eta-2)}{1 + \eta}T$. This linear behavior for different values of entropy  $S<S_{c}$, $S=S_{c}$, and $S>S_{c}$  is illustrated in  Fig.~\ref{fig:40}.

\begin{figure}[h]
	\begin{subfigure}{0.45\textwidth}\includegraphics[width=\textwidth]{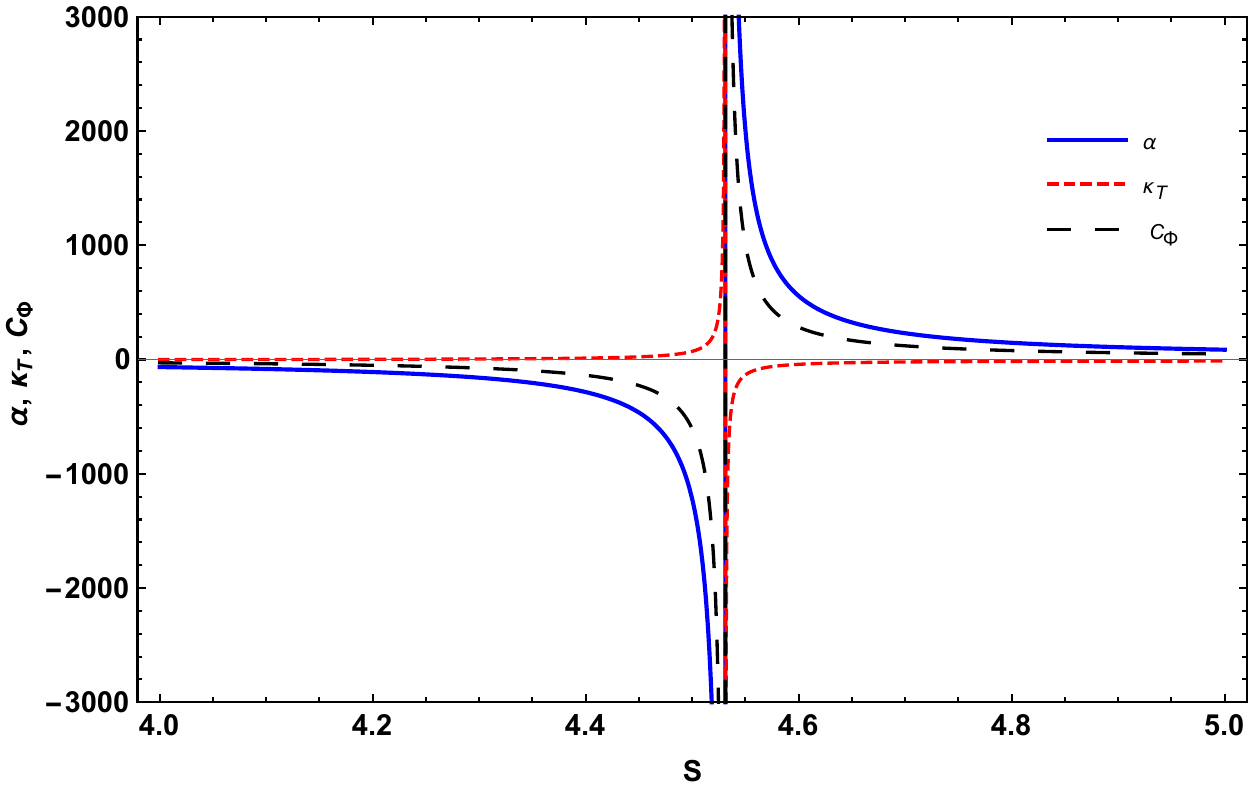}
		\caption{}
		\label{Al-1}
	\end{subfigure}
\begin{subfigure}{0.45\textwidth}\includegraphics[width=\textwidth]{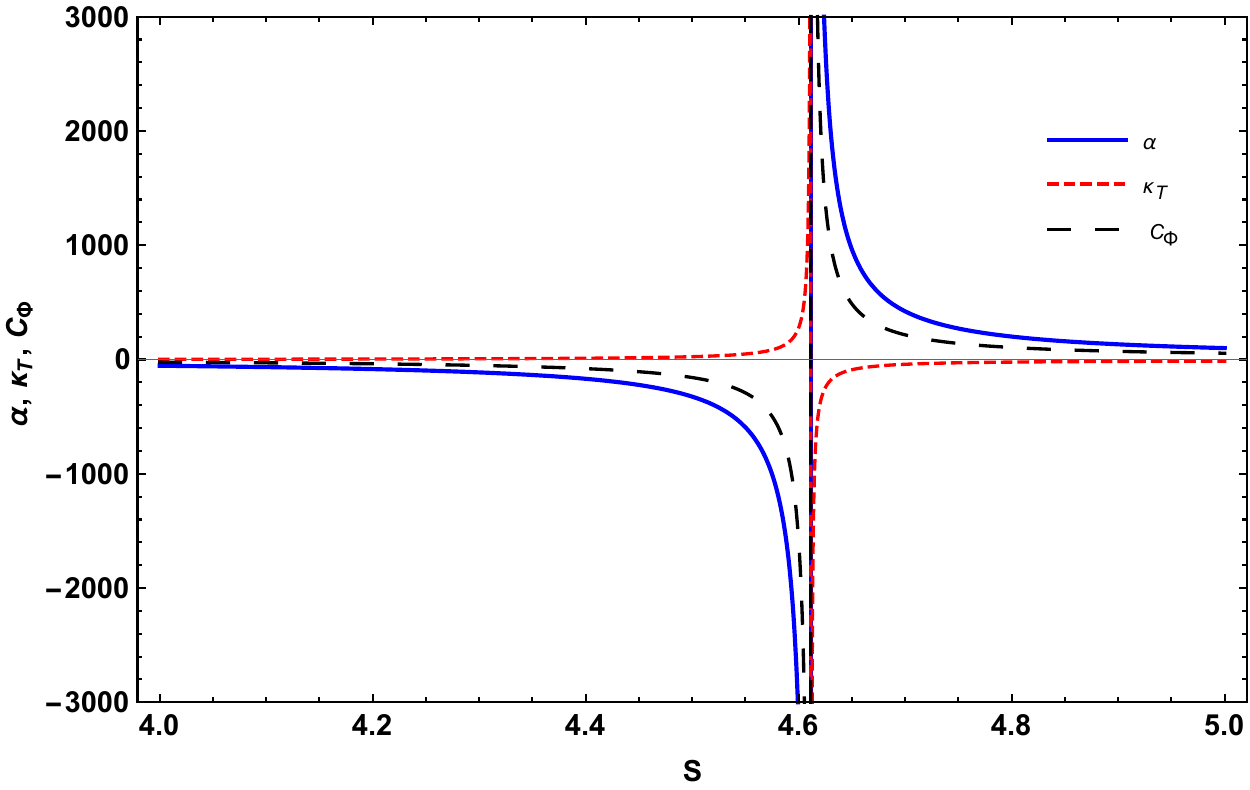}
	\caption{}
	\label{A2-1}
\end{subfigure}
	\begin{subfigure}{0.45\textwidth}\includegraphics[width=\textwidth]{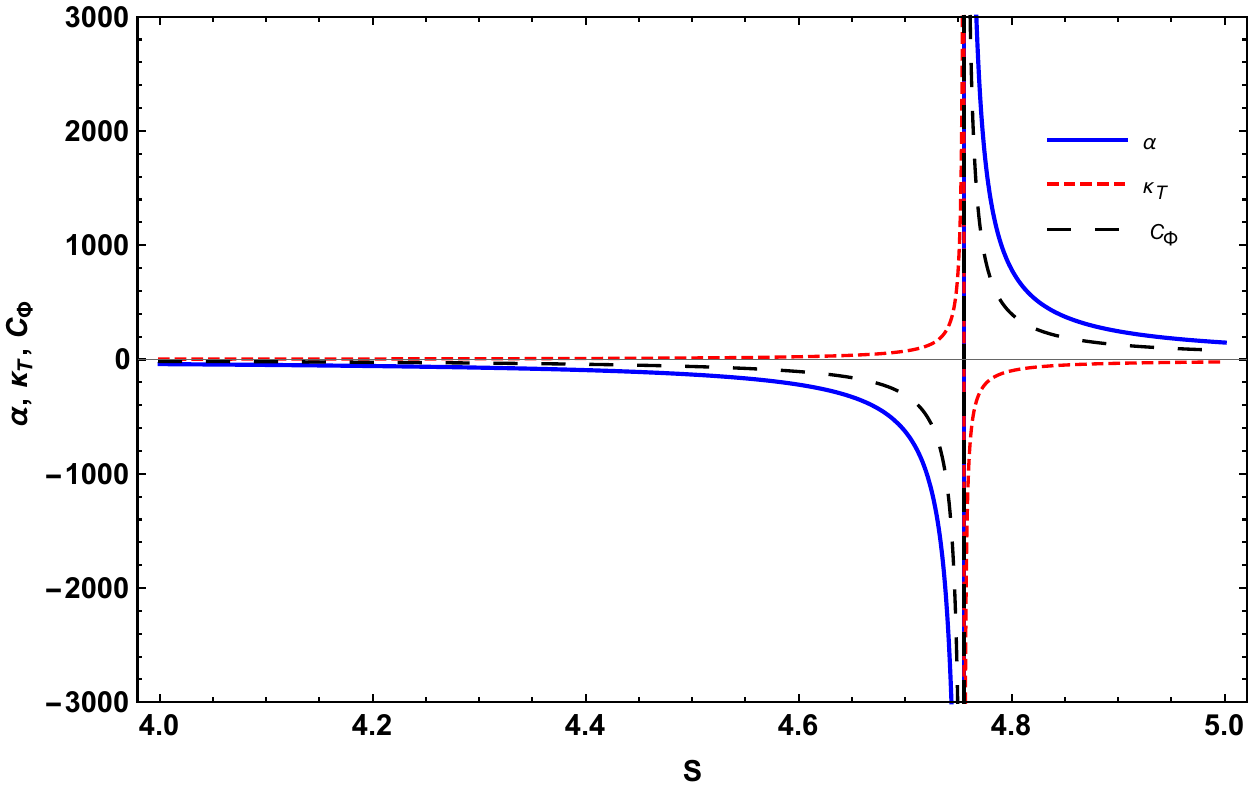}
		\caption{}
		\label{Ka-2}
	\end{subfigure}
	\caption{ Volume expansion coefficient, isothermal compressibility coefficient, and heat capacity at constant chemical potential vs. entropy for $a_0=1$, $\xi=3$ and $\eta=1$ when (\ref{Al-1})  $Q^2=.5 \,\,Q^2_c\,$, (\ref{A2-1})  $Q^2= \,\,Q^2_c\,$, and (\ref{Ka-2})  $Q^2=2 \,\,Q^2_c\,$.}
	\label{Akapa}
\end{figure}

There are of course other interesting response functions for these black holes that can be considered. Namely, the analog of volume expansion and isothermal compressibility coefficients, which are respectively denoted by $\alpha$ and $\kappa_{T}$ and can be computed from the following definitions
\begin{eqnarray}\label{alpha}
\alpha =\frac{1}{Q}{{\left( \frac{\partial Q}{\partial T} \right)}_{  \Phi }}=\frac{(2 \pi)^{2 + \xi} S^{\eta + \xi} ( \xi-1)}{(2 \pi)^{\xi} S^{1 + \eta + \xi} -  2^{\eta + \xi} a_0 \pi^{1 + \eta + \xi} S^{\xi} \eta -  2^{2 \xi} \pi^{1 + 2 \xi} Q^2 S^{\eta} ( \xi-2)}\,,
\end{eqnarray}
and
\begin{eqnarray}\label{kappa}
\kappa_{T}=-\frac{1}{Q}{{\left( \frac{\partial Q}{\partial   \Phi } \right)}_{T}}=\frac{2 \pi S^{ \xi-1} ( \xi-1) (S^{1 + \eta + \xi} -  2^{\eta} a_0 \pi^{1 + \eta} S^{\xi} \eta + 2^{\xi} \pi^{1 + \xi} Q^2 S^{\eta} \xi)}{Q \bigl(- (2 \pi)^{\xi} S^{1 + \eta + \xi} + 2^{\eta + \xi} a_0 \pi^{1 + \eta + \xi} S^{\xi} \eta + 2^{2 \xi} \pi^{1 + 2 \xi} Q^2 S^{\eta} ( \xi-2)\bigr)}\,.
\end{eqnarray}
By examining Eqs.~(\ref{alpha}) and (\ref{kappa}) alongside Eq.~(\ref{cphi}), it can be observed that both of the $\alpha$ and $\kappa_{T}$ share the same singularity as well as the heat capacity at constant chemical potential. For example, we have plotted these quantities as functions of entropy in three panels in Fig.~\ref{Akapa} for different values of the charge when $a_0=1$, $\eta=1$, and $\xi=3$. It is obvious from the figures that the value of the critical point is enlarged by increasing the charge. The Ehrenfest's equations can be solved at these diverging points to determine the order and the nature of the phase transitions. We address this point in the next subsection. We have also check out that the obtained heat capacities in this section satisfy the following relation that has been proposed in Ref.~\cite{Grumiller:2007ju} as
\begin{eqnarray}
{{C}_{  \Phi}}-{{C}_{Q}}=-\frac{TQ{{\alpha }^{2}}}{\kappa_{T}}\,.
\end{eqnarray}

\subsection{Ehrenfest’s Equations}
In general thermodynamics, the Gibbs free energy is defined by a Legendre transformation as $G\equiv H-TS$ where $H$ is the enthalpy of the system and its differential form is given by $dG=VdP-SdT$. That is the entropy and volume are given by derivative of $G$ with respect to the temperature and pressure, respectively. In this regard, if these quantities change discontinuously then the system undergo a first-order phase transition, in contrast, in a second-order phase transition the entropy and volume are continuous. Instead, the discontinuity in heat capacity suggests that there may be a second-order phase transition.
Ehrenfest defines a phase transition for two state functions $G_1$ and $G_2$ to be an $n$th-order transition if, at the transition point, $\partial^{n} G_1/\partial T^{n}\neq \partial^{n} G_2/\partial T^{n}$ and $\partial^{n} G_1/\partial P^{n}\neq \partial^{n} G_2/\partial P^{n}$ whereas all lower derivatives are equal. A well-known example is the second-order transition. However, in the case of charged black holes of Mass $M$, chemical potential $\Phi$, and charge $Q$ the Gibbs free energy is defined by
 \begin{eqnarray}
G\equiv M-\Phi Q-TS,
\end{eqnarray}
where according to the first low of black hole thermodynamics, i.e. $dM= T dS +\Phi dQ$, its variation becomes
\begin{eqnarray} \label{gfe}
dG= -Q d\Phi -S dT.
\end{eqnarray}
In a similar manner, when the entropy is continuous and the heat capacity is discontinuous,  to understand a phase transition phenomena one can use the standard Ehrenfest’s
prescription. Because of the differences between conventional systems and black holes, care must be exercised in applying Ehrenfest’s scheme to study phase transitions in black holes.

As depicted in Fig.~\ref{fig:1} the Hawking temperature is continuous when plotted in terms of $S$ for a fixed value of $Q$, i.e. isocharge curves, while in Fig.~\ref{fig:2} the heat capacity $C_Q$ has discontinuity, therefore there is no first-order phase transition for these black holes. Now we should investigate two Ehrenfest-like equations to study phase transition. A similar prescription in the case of Kerr black holes has been proposed in Ref.~\cite{Banerjee:2010bx}. In the second-order phase transition Gibbs free energy, entropy and charge are all continuous, then at the phase transition point one should has
\begin{eqnarray}\label{spt}
G_1=G_2,\qquad S_1=S_2, \qquad Q_1=Q_2.
\end{eqnarray}
Due to the continuity of Gibbs free energy in Eq.~(\ref{spt}), i.e. $dG=0$, and from Eq.~(\ref{gfe}) we obtain the following Maxwell relation
 \begin{eqnarray}\label{mrel}
 \left(\frac{\partial Q}{\partial T}\right)_{\Phi}=\left(\frac{\partial S}{\partial \Phi}\right)_{T}.
  \end{eqnarray}

Assume an infinitesimal increase in temperature and chemical potential which leads to $S_1+dS_1=S_2+dS_2$, then from (\ref{spt}) we find
 \begin{eqnarray}\label{seq}
dS_1=dS_2,
\end{eqnarray}
where the differential of entropy $S(T,\Phi)$ for each phase state is given by
  \begin{eqnarray}\label{}
dS= \left(\frac{\partial S}{\partial T}\right)_{\Phi} dT+\left(\frac{\partial S}{\partial \Phi}\right)_{T} d\Phi.
\end{eqnarray}
Substituting the relation (\ref{mrel}) into this variation, one arrive at
 \begin{eqnarray}\label{vs}
dS=\frac{C_{\Phi}}{T}\,dT+Q\,\alpha d\Phi,
\end{eqnarray}
 where $C_{\Phi}$ and $\alpha$ are given by Eqs.~(\ref{cphi}) and (\ref{alpha}), respectively. Now, we evaluate both side of relation (\ref{seq}) by this variation and then using the condition (\ref{spt}) for charges, we obtain
 \begin{eqnarray}\label{}
 \frac{{C_{\Phi}}_1}{T}\,dT+Q\,\alpha_1 d\Phi= \frac{{C_{\Phi}}_2}{T}\,dT+Q\,\alpha_2 d\Phi,
 \end{eqnarray}
 which yields one of the Ehrenfest's equations as
 \begin{eqnarray}\label{eh1}
 \left(\frac{\partial \Phi}{\partial T}\right)_S=-\frac{{C_{\Phi}}_1-{C_{\Phi}}_2}{Q T (\alpha_1-\alpha_2)}.
 \end{eqnarray}

Once again consider an infinitesimal change in temperature and chemical potential, the constancy of charge gives
  \begin{eqnarray}\label{qeq}
dQ_1=dQ_2.
\end{eqnarray}
 For variation of charge $Q(T,\Phi)$ one has
 \begin{eqnarray}\label{vq}
dQ= \left(\frac{\partial Q}{\partial T}\right)_{\Phi} dT+\left(\frac{\partial Q}{\partial \Phi}\right)_{T} d\Phi,
\end{eqnarray}
so that
 \begin{eqnarray}\label{}
dQ=Q\,\alpha\,dT-Q\,\kappa_{T} d\Phi,
\end{eqnarray}
 where $\kappa_{T}$ is given by Eq.~(\ref{kappa}). Similar to previous case, owing to the charges condition in (\ref{spt}), the equality (\ref{qeq}) for the two phases gives
the other Ehrenfest's equation as
 \begin{eqnarray}\label{eh2}
 \left(\frac{\partial \Phi}{\partial T}\right)_Q=\frac{\alpha_1-\alpha_2}{{\kappa_{T}}_1-{\kappa_{T}}_2}.
 \end{eqnarray}

In fact, for any true second-order phase transition both Ehrenfest's equations in (\ref{eh1}) and (\ref{eh2}) must be satisfied at the critical point. In order to investigate this fact, we have computed both the left hand side (LHS) and right hand side (RHS) of Ehrenfest's equations for infinitesimally close to the critical points and the results for different values of parameter space are given in Tab.~(\ref{tab2}). Consequently, the equivalence of the LHS and RHS of these equations for different values of couplings and black hole charge in dJT gravity confirms that we have the conventional \textit{second-order phase transition} for charged black holes in this theory.

 \begin{table}[h!]
\centering
\begin{tabular}{|c|c|c|c|c|c|c|}\hline
parameters & critical point & $Q^2$ & LHS (\ref{eh1}) & RHS (\ref{eh2}) & LHS (\ref{eh2}) & RHS (\ref{eh2}) \TBstrut\\\hline
  & $4.531$ & $.5 {Q_c}^2$ & $-15.695$ & $-15.695$ & $-15.695$ & $-15.695$  \TBstrut\\
$\eta=1$, $\xi=3$  & $4.611$ & ${Q_c}^2$ & $-11.295$ & $-11.295$ & $-11.295$ & $-11.2959$  \\
  & $4.755$ & $2 {Q_c}^2$ & $-8.236$ & $-8.236$ & $-8.236$ & $-8.236$  \\  \hline
  & $6.440$ & $.5 {Q_c}^2$ & $-8.114$ & $-8.114$ & $-8.114$ & $-8.114$  \TBstrut\\
$\eta=2$, $\xi=3$  & $6.583$ & ${Q_c}^2$ & $-5.865$ & $-5.865$ & $-5.865$ & $-5.865$  \\
  & $6.838$ & $2 {Q_c}^2$ & $-4.308$ & $-4.308$ & $-4.308$ & $-4.308$\\  \hline
   & $6.441$ & $.5 {Q_c}^2$ & $-7.524$ & $-7.524$ & $-7.524$ & $-7.524$  \TBstrut\\
$\eta=2$, $\xi=4$  & $6.579$ & ${Q_c}^2$ & $-5.434$ & $-5.434$ & $-5.434$ & $-5.434$  \\
  & $6.816$ & $2 {Q_c}^2$ & $-3.981$ & $-3.981$ & $-3.981$ & $-3.981$ \\ \hline
\end{tabular}
 \caption[]{Ehrenfest's equations (\ref{eh1}) and (\ref{eh2}) when evaluated close to the critical values of entropy for $a_0=1$ and given parameters.}
\label{tab2}
\end{table}
\section{Thermodynamic Geometry}\label{4}

We use geometric methods to study the thermodynamics of the charged black holes in dJT gravity and their phase transitions. These methods involve defining a metric on the thermodynamic phase space and computing the Riemann scalar curvature. The curvature can reveal important information about the criticality and microstructures of the black hole system\cite{Aman:2003ug,Sarkar:2006tg,Mirza:2007ev,Ruppeiner:1995zz}. We investigate how the curvature reflects the divergence of the heat capacities that we found in the previous section. We also compare different metrics, such as the Weinhold, Ruppeiner, and new thermodynamic geometry (NTG) metrics.
\subsection{Weinhold Geometry}
The Weinhold metric is a way of measuring how the internal energy of a system changes with respect to its entropy and other extensive parameters, such as volume, charge and number of particles. It defines a Riemannian metric on the thermodynamic phase space, which captures the intrinsic geometry of the system. In the black hole thermodynamics, the energy of the system is described by the black hole mass, thus Weinhold metric is defined as follows
\begin{equation}\label{eq-Smarr1}
ds_{W}^2= g_{ab} ^W \,dX^a dX^b , \qquad  g_{ab} ^W =\frac{\partial^2 M}{\partial X^a \partial X^b}\,,
\end{equation}
where the 2D surface in the phase space is specified by coordinates $X^a=(S,Q)$. We use Eq.~(\ref{eq-Smarr1}) to calculate the Weinhold curvature on this surface given by
\begin{equation}\label{WC}
R_{W}=\frac{2 \pi^2 S^{\eta + 2 \xi} \Bigl(a_0 \pi \eta \bigl(2^{1 + \eta} \pi^{\eta} + (2 \pi)^{\eta} (\eta -  \xi)\bigr) + S^{1 + \eta} (\xi-1)\Bigr) (\xi-1)}{S \bigl(- S^{1 + \eta + \xi} + 2^{\eta} a_0 \pi^{1 + \eta} S^{\xi} \eta + 2^{\xi} \pi^{1 + \xi} Q^2 S^{\eta} ( \xi-2)\bigr)^2}\,.
\end{equation}

The behavior of the Weinhold curvature as a function of entropy for a given value of the charge Q is given by a solid blue line in Fig.~\ref{fig:4}.  As is observed, the critical behavior of $R_W$ is coincident with the one of $C_\Phi$ which is given by red dashed line. One can also easily check this fact by comparing the denominators of Eqs.~(\ref{cphi}) and (\ref{WC}) which occurs at the same temperature $T'_c$ which is explored in Fig.~\ref{fig:5}. Therefore, the Weinhold curvature can be written as
\begin{equation}\label{wt}
R_{W}=\frac{Y}{16 \pi^4 S^{2 \eta} ( \xi-2)^2 \, ( T - T'_{c})^2}\,,
\end{equation}
where the function $Y=2 \pi^2 S^{\eta} (\xi-1) \Bigl(a_0 \pi \eta \bigl(2^{1 + \eta} \pi^{\eta} + (2 \pi)^{\eta} (\eta -  \xi)\bigr) + S^{1 + \eta} (\xi-1)\Bigr) $ is regular in Eq.~(\ref{wt}). The Weinhold curvature in Eq.~(\ref{wt}) follows the power-law $R_{W}\sim {(T - T'_{c} )}^{-\alpha}$ with $\alpha=2$ while in the case of $C_\Phi$ it was $\alpha=1$.
\subsection{Ruppeiner  Geometry}

To study the microscopic properties of 2D JT gravity, we use Ruppeiner geometry, which is based on the entropy as the main thermodynamic variable. This allows us to extend this approach to black hole systems, which have entropy and charge as the natural coordinates for fluctuations in the parameter space. In fact, the curvature of the Ruppeiner geometry indicates a correlation length which has some singularities at critical points. The Ruppeiner metric is defined by $g_{ab}^{Ru}=-\frac{\partial^2 S}{\partial X^a \partial X^b}$ and according to the black hole thermodynamics, it can be written as the following line element
\begin{equation}\label{eq-Smarr}
ds^2_{Ru}=-\,T^{-1}\,  g_{ab} ^W \,dX^a dX^b \,,
\end{equation}
where the 2D surface is described by coordinates $X^a=(S,Q)$. This relation shows that the Ruppeiner metric and the Weinhold metric are related by a conformal factor $T^{-1}$, where $T$ is the temperature of the black hole. Employing the physical parameters of the dJT black hole in Eq.~(\ref{TM}), we compute the Ruppeiner metric as follows
\begin{equation}\label{RRu}
R_{Ru}=\frac{F(S,Q)}{A(S,Q) \,\,B(S,Q)^2 }\,,
\end{equation}
where the functions $F$, $A$ and $B$ are given by:
\begin{eqnarray}
F&=&2^{-1- 2 \xi} \pi^{-2 \xi} S^{ \xi-1} \Biggl(2^{3 \eta + 2 \xi} a_0^3 \pi^{3 + 3 \eta + 2 \xi} S^{2 \xi} \eta (1 + \eta -  \xi) ( \eta + \xi-2)\\
&& -  S^{1 + 3 \eta} (\xi-3) \Bigl(- (2 \pi)^{2 \xi} S^{2 + 2 \xi} \xi -  8^{\xi} \pi^{1 + 3 \xi} Q^2 S^{1 + \xi} (3 + \xi) + 16^{\xi} \pi^{2 + 4 \xi} Q^4 \bigl( 2 \xi ( \xi-1) -3 \bigr)\Bigr)\nonumber\\
&& + 4^{\eta + \xi} a_0^2 \pi^{2 (1 + \eta + \xi)} S^{\eta + \xi} \biggl(2^{\xi} \pi^{1 + \xi} Q^2 \eta \bigl((\eta-2)^2 -  \xi^2\bigr) + S^{1 + \xi} \Bigl(1 -  \eta \bigl(5 + \eta (7 + \eta)\bigr) - 2 \xi + \eta (8 + \eta) \xi + (1 - 2 \eta) \xi^2\Bigr)\biggr)\nonumber\\
&& + 2^{\eta + 2 \xi} a_0 \pi^{1 + \eta + 2 \xi} S^{2 \eta} \Bigl(S^{2 + 2 \xi} \bigl(3 + 3 \eta - 5 \xi + \eta (5 + \eta) \xi -  (\eta-2) \xi^2\bigr)\nonumber\\
&& -  2^{\xi} \pi^{1 + \xi} Q^2 S^{1 + \xi} \bigl(-3 \eta (7 + \eta) + 3 (1 + \eta)^2 \xi + (\eta-1) \xi^2\bigr) + 4^{\xi} \pi^{2 + 2 \xi} Q^4 ( \eta + \xi-2) \bigl(-2 ( \xi-2) \xi + \eta ( 2 \xi-3)\bigr)\Bigr)\Biggr)\,,\nonumber
\\\nonumber
A&=& 2^{\xi} \pi^{1 + \xi} Q^2 S^{\eta} - 2^{\eta} a_0 \pi^{1 + \eta} S^{\xi} - S^{1 + \eta + \xi}\,,\nonumber\\
B&=&S^{\xi} ( 2^{\eta} a_0 \pi^{1 + \eta} \eta - S^{1 + \eta}) + 2^{\xi} \pi^{1 + \xi} ( \xi -2)  Q^2 S^{\eta}\,.\nonumber
\end{eqnarray}

This thermodynamic curvature has singularities when $A=0$ or $B=0$. When the temperature is zero $(T=0)$ and the charge is equal to the extremal charge $Q^2=Q_E^2$, we have $A=0$, thus the curvature diverges in the extremal limit. Another singularity occurs when $B=0$ at $Q^2=Q_c^2$. The temperature corresponding to this charge was denoted by $T_c$. To look for our goal in this paper, we can recast the thermodynamic curvature in Eq.~(\ref{RRu}) in terms of $S$ and $T$ by substituting from Eq.~(\ref{eq-TM}). The result is
\begin{equation}
R_{Ru}(S,T)=\frac{H(S,T)}{T \Bigl(S^{\eta} \bigl( 4 \pi^2 T ( \xi-2)- S (\xi-3) \bigr) + a_0 \pi \bigl(2^{1 + \eta} \pi^{\eta} -  (2 \pi)^{\eta} (\eta + \xi)\bigr)\Bigr)^2}\,,
\end{equation}
where again $H(S,T)$ is a lengthy function of the temperature and entropy. Therefore, we can rewrite  the thermodynamic curvature in terms of the critical temperature as follows
\begin{equation}
R_{Ru}(S,T)= \frac{H(S,T)}{16 \pi^4 \, ( \xi-2)^2\, S^{2 \eta}\, T\, ( T-T'_{c})^2 }\,.
\end{equation}
The thermodynamic curvature has a power-law behavior near the critical temperature, given by $R_{Ru}\sim {(T  - T'_{c}  )} ^{-2}$. However, this does not apply when $\xi=2$, as there is no critical temperature for the thermodynamic curvature in this case. Instead, the thermodynamic curvature shows a linear behavior similar to the heat capacity at constant chemical potential in Eq.~(\ref{Tcp2}) when $\xi=2$.

\begin{figure}[h!]
	\begin{subfigure}{0.45\textwidth}\includegraphics[width=\textwidth]{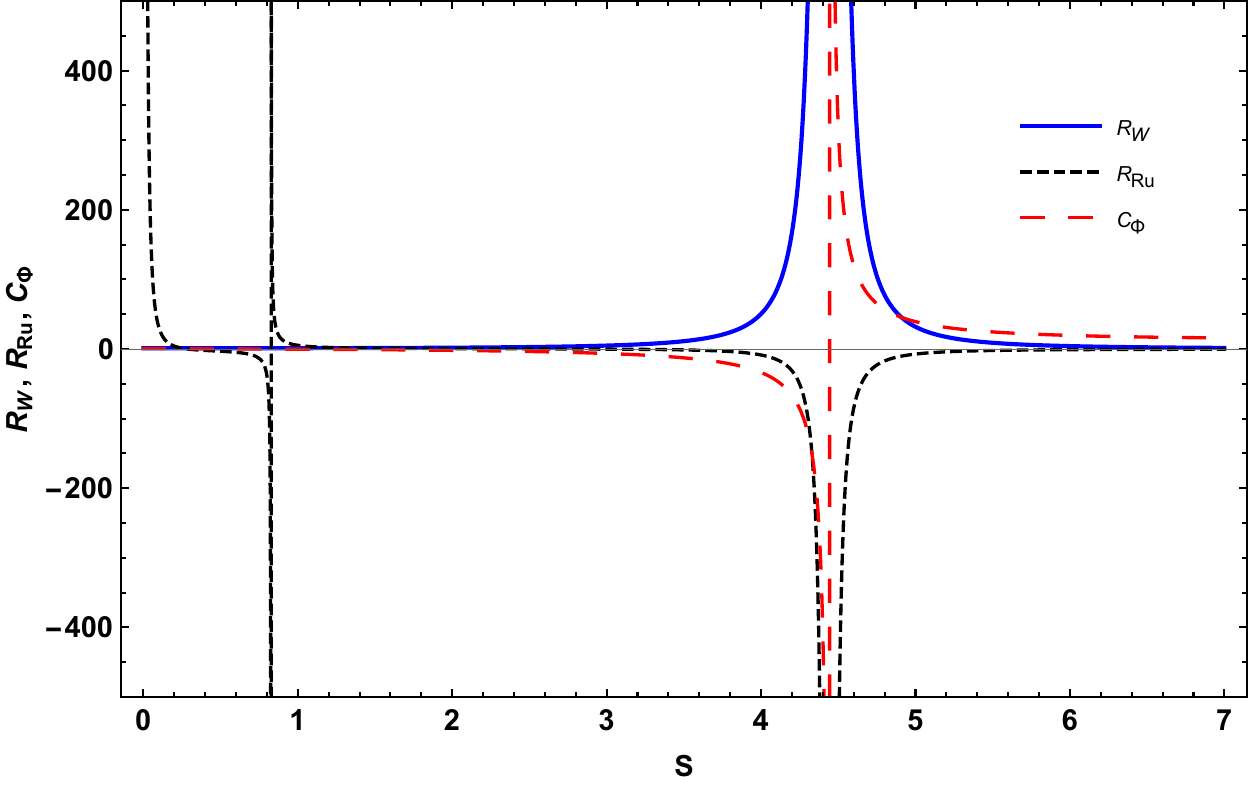}
		\caption{}
		\label{fig:4-1}
	\end{subfigure}
	\begin{subfigure}{0.45\textwidth}\includegraphics[width=\textwidth]{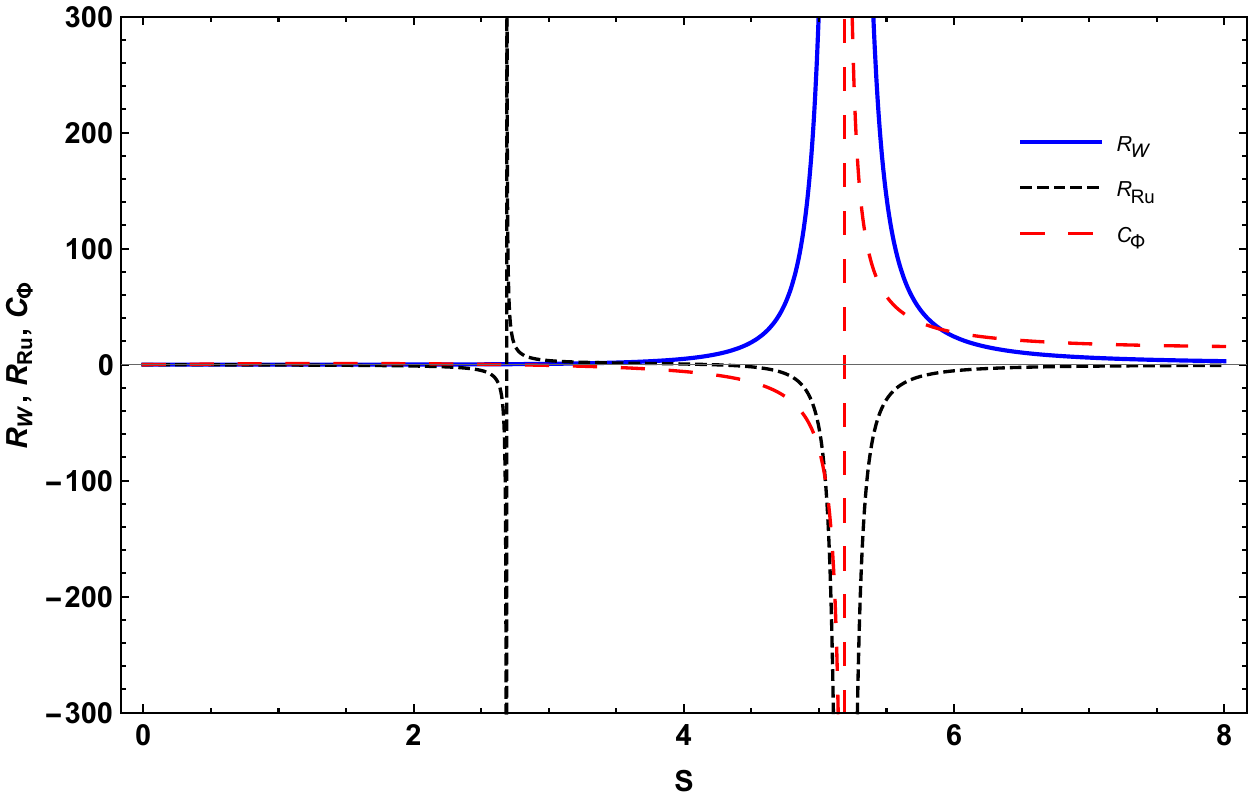}
		\caption{}
		\label{fig:4-2}
	\end{subfigure}
	\caption{$R_{Ru}$, $R_{W}$ and $C_{\Phi}$ vs. entropy when $a_0=1$ and $\eta=1;$ (\ref{fig:4-1}) $\xi=2$  $Q=0$  and  (\ref{fig:4-2})  $\xi=3$ and  $Q=\tfrac{1}{2 }$.}
	\label{fig:4}
\end{figure}

We show the Ruppeiner and Weinhold curvatures, as well as the heat capacity at constant chemical potential ($C_{\Phi}$), as functions of entropy in Fig.~\ref{fig:4}. All three quantities have the same divergence pattern. A remarkable feature of the case $\xi=2$ is that the singular values of the $R_{Ru}$, $R_{W}$, and $C_{\Phi}$ are independent of $Q^2$.

\begin{figure}[h!]
	\begin{subfigure}{0.45\textwidth}\includegraphics[width=\textwidth]{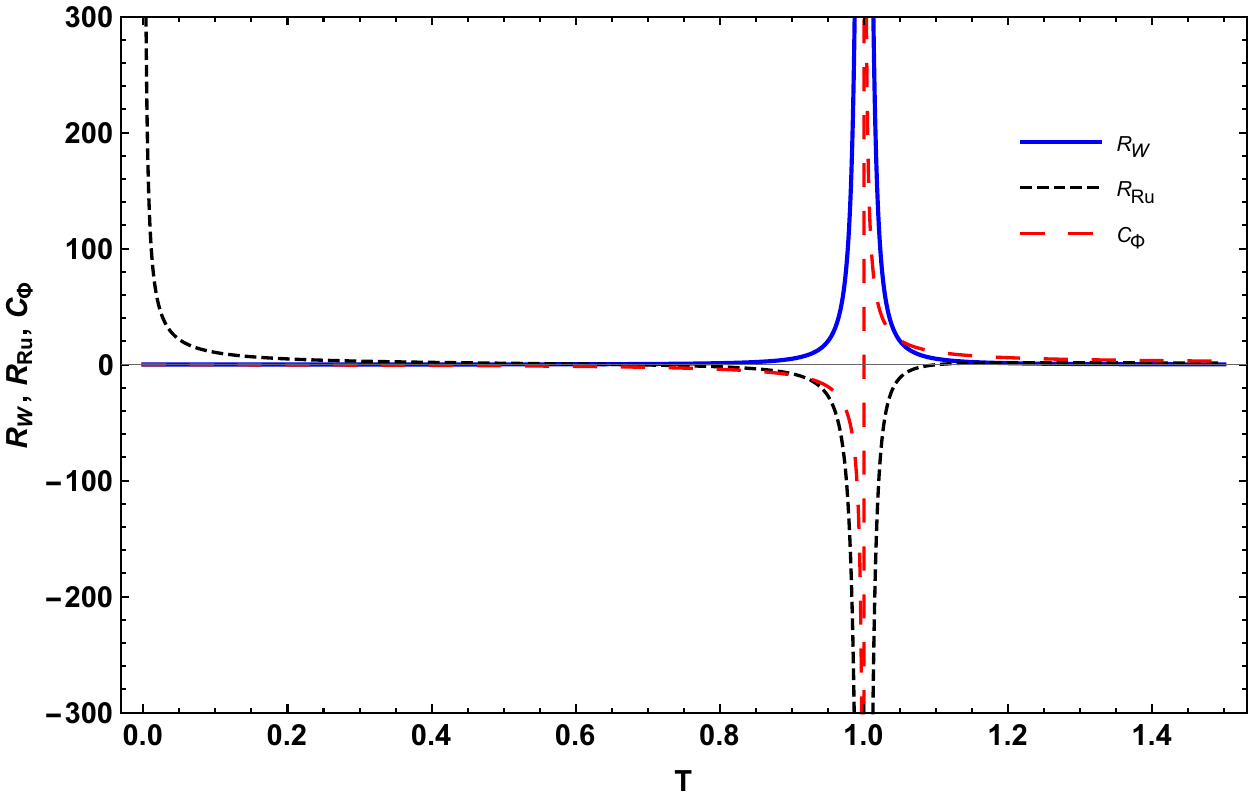}
		\caption{}
		\label{fig:5-1}
	\end{subfigure}
	\begin{subfigure}{0.45\textwidth}\includegraphics[width=\textwidth]{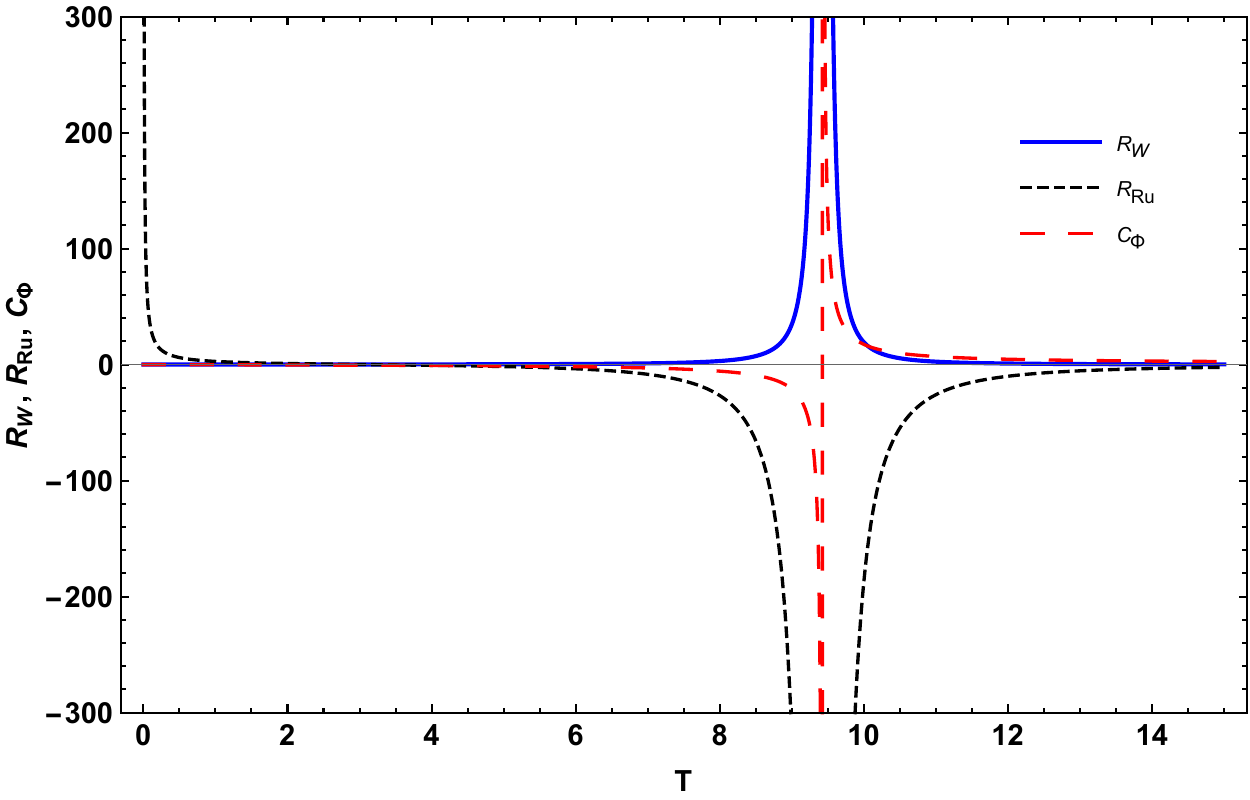}
		\caption{}
		\label{fig:5-2}
	\end{subfigure}
	\caption{$R_{Ru}$, $R_{W}$ and $C_{\Phi}$ vs. temperature when $a_0=1$,  $S=1;$  (\ref{fig:5-1}) $\eta=1,$ $\xi=3$  and (\ref{fig:5-2})  $\eta=2,$  $\xi=3$.}
	\label{fig:5}
\end{figure}
Also, Fig.~\ref{fig:5} shows how $R_{Ru}$, $R_{W}$, and $C_{\Phi}$ vary with temperature. All three quantities have the same divergence behavior at $T'_{c}$. We also note that the Ruppeiner curvature, $R_{Ru}$, diverges at $T=0$.
\subsection{New Thermodynamic Geometry}
It has been proposed in \cite{HosseiniMansoori:2019jcs} that there is a new formulation for Ruppeiner metric in which the thermodynamic potential $\Xi$ is related to the mass by a Legendre transformation and represents a one-to-one correspondence between critical points of the heat capacities and those of the thermodynamic curvatures. In this respect, when $\Xi=M$ the curvature of this metric diverges at critical points of $C_Q$ and when $\Xi=M-\Phi Q$ the singularities correspond to phase transitions of $C_\Phi$. The metric of NTG is given by
\begin{equation}\label{Ru1}
ds_{NTG}^2=\frac{1}{T}\left( \eta_i^{ j} \, \frac{\partial^2\Xi}{\partial X^j
	\partial X^l} \, d X^i d X^l \right)\,,
\end{equation}
where  $\eta_i^{ j}$ takes the form of a diagonal matrix with entries $(-1,1,...,1)$ and the $X^{i}$ may refer to either intensive or extensive variables depending to the kind of ensemble. A noteworthy aspect of this framework is its ability to elucidate the connection between phase transitions and curvature singularities on a one-to-one basis. To illustrate, if we select $\Xi= G( T,  \Phi)$ where $ G=M- T S- \Phi  Q$ is the Gibbs free energy and $X^{i}=( T, \Phi)$ within the NTG metric, this correspondence becomes apparent \cite{HosseiniMansoori:2019jcs}.

From Eq.~(\eqref{Ru1}), we have
\begin{equation}\label{CFTmetric_2}
g^{NTG}_{ F}=\frac{1}{ T} \, \text{diag}\Big(-\frac{\partial^2  G}{\partial  T^2 },\frac{\partial^2  G}{\partial  \Phi^2 }\Big)=C_{ \Phi} \, \text{diag}\Big( T^{-2},-\mathcal{F}\Big)=C_{ \Phi}\, \hat{g}, \hspace{0.25cm} \text{with} \hspace{0.25cm} \mathcal{F}=\frac{\Big(\frac{\partial  Q}{\partial  \Phi}\Big)_{ T}}{ T C_{ \Phi}}\,,
\end{equation}
in which we have used the first law in the variation of the free energy, i.e. $dG=-Sd T- Qd  \Phi$. Another represetation that can be used for the above metric is given by \begin{equation}\label{met1}
g^{NTG}_{F}=\frac{1}{T} \left( \begin{matrix}
\frac{C_{ \Phi}}{ T}& 0\\
0 & - \Big(\frac{\partial  Q}{\partial  \Phi}\Big)_{ T}\\
\end{matrix}\right)\,,
\end{equation}
where $C_{ \Phi}$ is the heat capacity at constant chemical potential. But from Eq.~(\ref{cphi}), this is explicitly given in terms of $S$ and $Q$ and one cannot write it as a function of $T$ and $\Phi$ just by using Eqs.~(\ref{eq-S}) and (\ref{TM}). However, this is possible only for some particular values of constant parameters. Due to this fact, in the rest of this section, we consider two families of the charged black hole solutions in the canonical ensemble with $\eta=1$ but for different values $\xi=2$ and $\xi=3$.
\subsubsection{NTG for $\eta=1$ and $\xi=2$.}
Solving the relation of temperature in Eq.~(\ref{TM}) for $\eta=1$ and $\xi=2$, one can find the entropy as a function of $T$ and $\Phi$ as follows
\begin{eqnarray}
S=\tfrac{1}{2} \Bigl(\Phi^2 \pi + 4 \pi^2 T - \sqrt{ (- \Phi^2 \pi - 4 \pi^2 T)^2-8 a_0 \pi^2 }\Bigr),
\end{eqnarray}
and the heat capacity at constant chemical potential given in Eq.~(\ref{cphi}) can be redefined as $C_{\Phi}(T,\Phi)$. Thus, the metric elements  (\ref{met1}) are given by
\begin{eqnarray}
g_{TT}&=&\frac{2 \pi^2 \Bigl(-\Phi^2 - 4 \pi T + \sqrt{ (\Phi^2 + 4 \pi T)^2-8 a_0 }\Bigr)}{T \sqrt{ (\Phi^2 + 4 \pi T)^2-8 a_0 }}\,,\\\nonumber
g_{\Phi \Phi}&=& -\frac{8 a_0 + (3 \Phi^2 + 4 \pi T) \Bigl(-\Phi^2 - 4 \pi T + \sqrt{ (\Phi^2 + 4 \pi T)^2-8 a_0 }\Bigr)}{4 T \sqrt{ (\Phi^2 + 4 \pi T)^2-8 a_0 }}\,,
\end{eqnarray}
and the thermodynamic curvature $R_{NTG}(T,\Phi)$ can be computed for this metric in the 2D surface $X^{i}=( T, \Phi)$ of the phase structure. In the first two parts we have shown that the singularities of the Ruppeiner and Weinhold curvatures are coincident with the critical points of heat capacity at constant chemical potential, but here we will show that  in the case of $R_{NTG}(T,\Phi)$ these points are correspond to ones of $C_Q$. Since the heat capacity $C_Q$ in Eq.~(\ref{eq17}) or (\ref{cq}) is a function of entropy, to compare the behaviors one needs to rewrite the curvature in terms of $S$. For example, by using Eqs.~(\ref{eq-S}) and (\ref{TM}), it becomes
\begin{equation} \label{rsq1}
R_{NTG}(S,Q)=\frac{2 \pi^2 (3 \pi Q^2 -  a_0 S) \bigl(4 a_0^2 \pi^4 S + 4 a_0 (2 \pi^5 Q^2 + 3 \pi^2 S^3) - 3 (4 \pi^3 Q^2 S^2 + S^5)\bigr)}{(4 \pi^3 Q^2 - 2 a_0 \pi^2 S -  S^3) (8 \pi^3 Q^2 - 2 a_0 \pi^2 S + S^3)^2}\,,
\end{equation}
or equivalently in terms of temperature as
\begin{equation} \label{rts1}
R_{NTG}(T,S)=- \frac{\bigl(4 a_0^2 \pi^4 - 3 S^3 (S - 2 \pi^2 T) + 4 a_0 \pi^2 S (S -  \pi^2 T)\bigr) \bigl(3 S^2 + 2 \pi^2 (a_0 - 6 S T)\bigr)}{256 \pi^6 S^4 T (T -  \hat{T}_c)^2}\,,
\end{equation}
where $\hat{T_c}$ is the critical temperature evaluated at the corresponding parameters, i.e. $T_{c}(S)\Big|_{\eta=1,\xi=2}=\hat{T}_c$. The expression in Eq.~(\ref{rts1}) asserts that the curvature singularities and phase transitions of the heat capacity at $\eta=1$ and $\xi=2$ are equivalent.
\subsubsection{NTG  for  $\eta=1$ and $\xi=3$.}
\begin{figure}[h!]
	\begin{subfigure}{0.45\textwidth}\includegraphics[width=\textwidth]{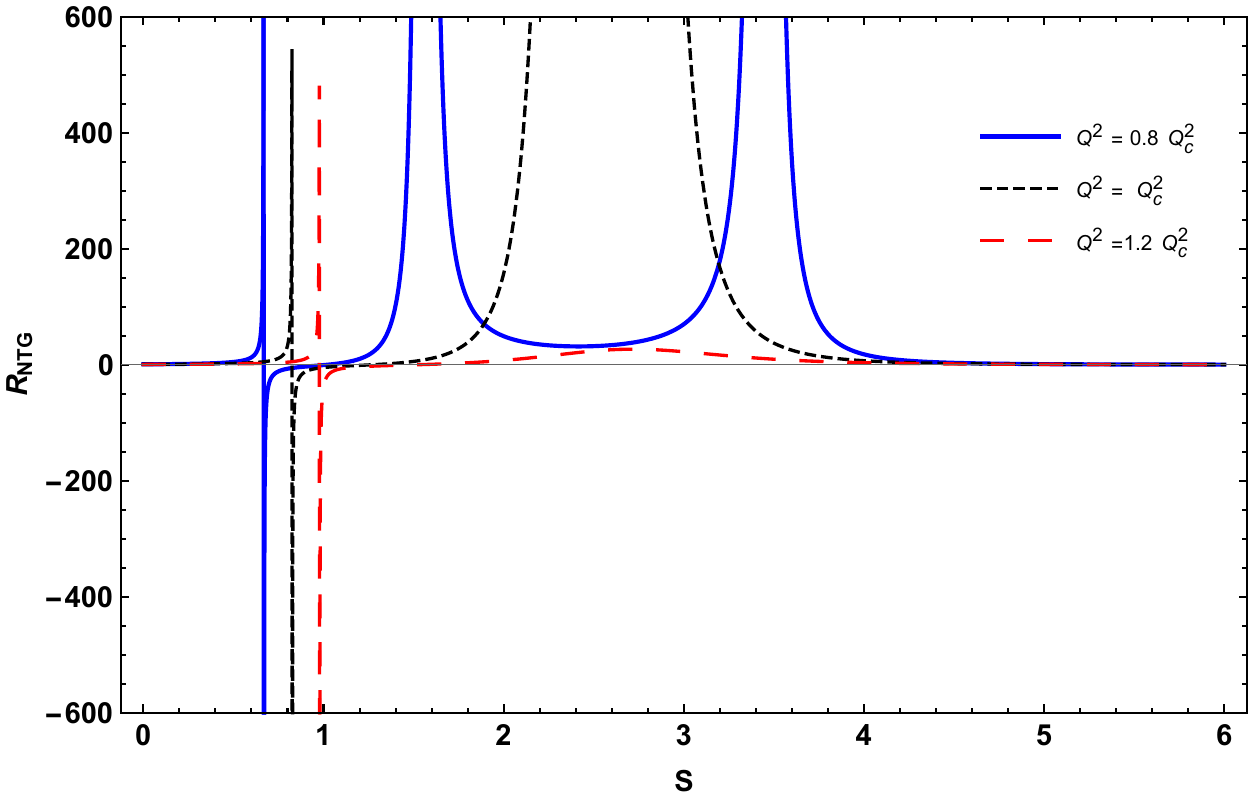}
		\caption{}
		\label{fig:6-1}
	\end{subfigure}
	\begin{subfigure}{0.45\textwidth}\includegraphics[width=\textwidth]{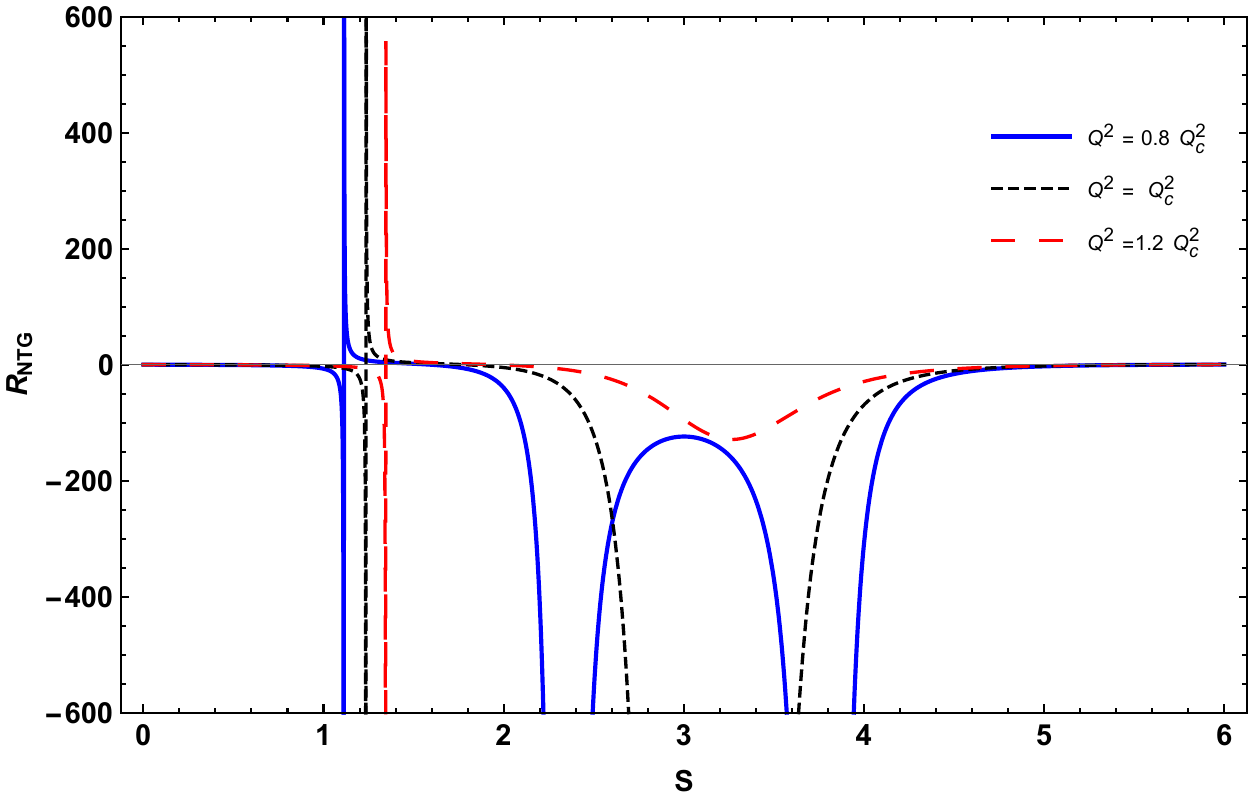}
		\caption{}
		\label{fig:6-2}
	\end{subfigure}
	\caption{The NTG curvature vs. entropy when $a_0=1$ and $\eta=1\,;$ (\ref{fig:6-1})  $\xi=2$  and   (\ref{fig:6-2}) $\xi=3$.}
	\label{fig:6}
\end{figure}

\begin{figure}[h!]
	
	\begin{subfigure}{0.45\textwidth}\includegraphics[width=\textwidth]{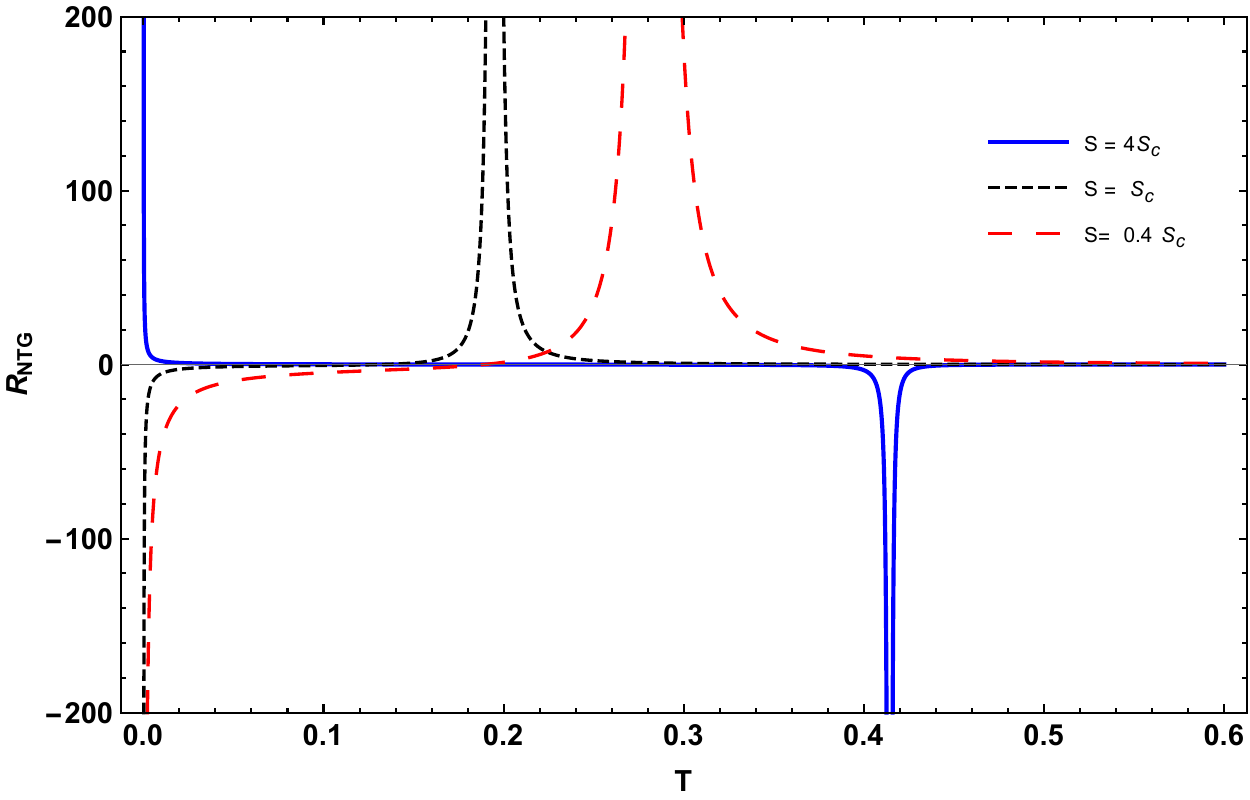}
		\caption{}
		\label{fig:7-1}
	\end{subfigure}
	\begin{subfigure}{0.45\textwidth}\includegraphics[width=\textwidth]{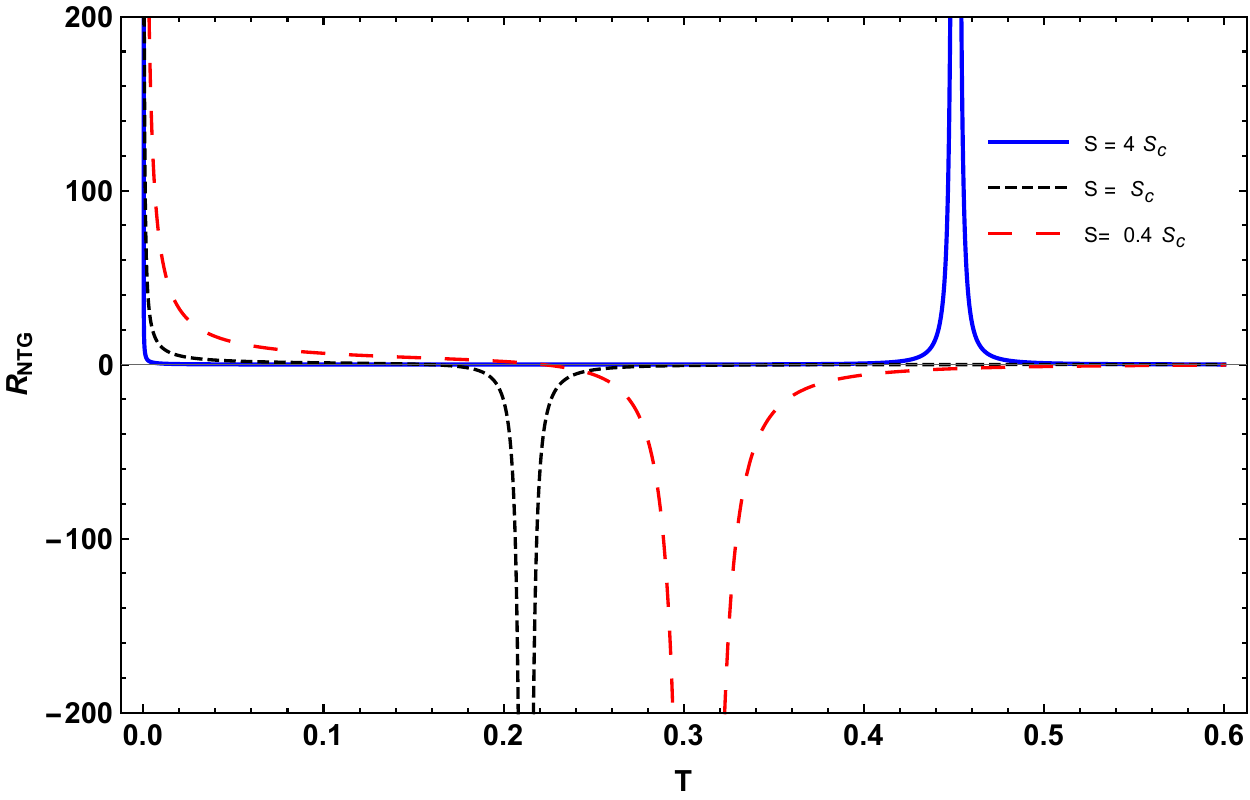}
		\caption{}
		\label{fig:7-2}
	\end{subfigure}
	\caption{The NTG curvature vs. temperature when $a_0=1$ and $\eta=1\,;$ (\ref{fig:7-1})  $\xi=2$  and   (\ref{fig:7-2}) $\xi=3$.}
	\label{fig:7}
	
\end{figure}

Following the same prescription in the previous part, we find the entropy in terms of $T$ and $\Phi$ for $\eta=1$ and $\xi=3$ as
\begin{eqnarray}
S=- \frac{2 \pi^2 T - \sqrt{2 (2 a_0 \Phi^2 \pi^2 + 2 \pi^4 T^2- a_0 \pi^2 )}}{ 2 \Phi^2-1}.
\end{eqnarray}
We can substitute this into the heat capacity at constant chemical potential in Eq.~(\ref{cphi}), and obtain the NTG metric (\ref{met1}) as
\begin{eqnarray}\label{eq44}
g_{TT}&=&\frac{2 \pi^2 }{ T ( 2 \Phi^2-1)} \Bigl( \frac{\pi T}{\sqrt{a_0 (  \Phi^2- \tfrac{1}{2}) + \pi^2 T^2}}-1 \Bigr)\,,\\\nonumber
g_{\Phi \Phi}&=&- \frac{a_0 - 4 a_0 \Phi^4 + (1 + 6 \Phi^2) \pi T \Bigl(-2 \pi T +  \sqrt{a_0 ( 4 \Phi^2-2) + 4 \pi^2 T^2}\Bigr)}{a_0 ( 2 \Phi^2-1) + \pi T \bigl(2 \pi T -  \sqrt{-2 a_0 + 4 a_0 \Phi^2 + 4 \pi^2 T^2}\bigr)}\\\nonumber
&\times&\frac{\Bigl(-2 \pi T + \sqrt{a_0 ( 4 \Phi^2-2) + 4 \pi^2 T^2}\Bigr)^2}{2 ( 2 \Phi^2-1)^3 T}\,.
\end{eqnarray}
Again the thermodynamic curvature $R_{NTG}(T,\Phi)$ can be written in terms of entropy and charge as
\begin{equation} \label{rsq2}
R_{NTG}(S,Q)=\frac{2 S \bigl(-12 a_0^3 \pi^6 S^4 - 20 a_0^2 \pi^4 S^6 + (-8 \pi^4 Q^2 S + S^5)^2 + a_0 \pi^2 (704 \pi^8 Q^4 + 144 \pi^4 Q^2 S^4 + 3 S^8)\bigr)}{(24 \pi^4 Q^2 - 2 a_0 \pi^2 S^2 + S^4)^2 (-8 \pi^4 Q^2 + 2 a_0 \pi^2 S^2 + S^4)}\,,
\end{equation}
 or in terms of entropy and temperature as
\begin{equation} \label{rts2}
R_{NTG}(T,S)=\frac{2 a_0 (a_0 \pi^2 + S^2)^2 - 11 a_0 \pi^2 S (a_0 \pi^2 + S^2) T + \pi^2 S^2 (11 a_0 \pi^2 + S^2) T^2}{18 \pi^4 S^4 T (T -  \tilde{T}_c)^2}.
\end{equation}
where $T_{c}(S)\Big|_{\eta=1,\xi=3}=\tilde{T}_{c}$. Now to accomplish our comparison, we have plotted $R_{NTG}(T,S)$ vs. entropy and temperature in Figs.~\ref{fig:6} and \ref{fig:7} respectively. The curves in Fig.~\ref{fig:6} are related to different values of charge $Q$ for the above parameter space, i.e. Fig.\ref{fig:6-1} for curvature in Eq.~(\ref{rsq1}) with $\xi=2$ and Fig.~\ref{fig:6-2} for curvature in Eq.~(\ref{rsq2}) with $\xi=3$, and three distinct cases of black holes are distinguished by them:

\begin{enumerate}
	\item  For $Q^2<Q_c^2$ by solid blue line, there are three divergent points for $R_{NTG}$. One of them corresponds to the $T=0$ from Eq.~(\ref{TM}) while the two others are coincident  with the divergent points of $C_Q$ depicted in Fig.~\ref{fig:2}.
	\item For $Q^2=Q_c^2$ by black dashed line, there are two divergent points for $R_{NTG}$, one is related to $T=0$ and again the other corresponds to the divergent point of $C_Q$.
	\item For $Q^2>Q_c^2$ by red dashed line, there is just a single divergent point  corresponds to the $T=0$.
\end{enumerate}

Also different behaviors of $R_{NTG}$ for $\xi=2$ and $\xi=3$ in Figs.~\ref{fig:6} have physical interpretation in the context of thermodynamics. For example, the positive values of it in Fig.~\ref{fig:6-1} indicates the attraction between interparticle interactions of a bosonic gas while the negative values in Fig.~\ref{fig:6-2} show a repulsive effect in a fermionic gas.

The critical behavior of $R_{NTG}(T,S)$ vs. temperature for different values of the entropy is illustrated in Fig.~\ref{fig:7}. Regardless of the trivial divergences at zero temperature, there is also a divergent point $\tilde{T_c}$ for each value of $S$ which is coincident with the critical temperature of $C_{Q}$ as shown in Fig.~\ref{fig:3}. Similarly, the curves of different $\xi$ have opposite signs with a special physical interpretation related to itself.

\section{Conclusion and outlook}\label{5}
In this paper we have investigated the thermodynamic properties of charged AdS black holes in the 2D dJT gravity which may be dual to a 1D SYK model from the holographic point of view. In the context of AdS black hole thermodynamics, the charged black hole has a van der Waals-Maxwell like phase structure. In particular, we computed the heat capacities at constant charge and chemical potential and the thermodynamic geometries in three different approaches. We showed that the corresponding heat capacities have the same power-law behaviors at their critical points given by the temperatures $(T_c, T'_c)$, that is $C_{Q}\sim {(T  - T_{c} )} ^{-1}$ and $C_{\Phi}\sim {(T - T'_{c} )} ^{-1}$. As shown in Fig.~\ref{fig:2}, the heat capacity $C_{Q}$ has different behaviors for $Q^2<Q_c^2$, $Q^2=Q_c^2$, and $Q^2>Q_c^2$ so that they distinguish three different types of black holes that describe two divergences, one divergence and no divergence, respectively. Moreover, it was shown that $C_{\Phi}$ has a linear behavior in terms of the temperature for specific value of $\xi=2$. We have also made a systematic study of the phase transition in the corresponding charged black holes by employing the Ehrenfest's scheme. We find a couple of points, dictated by the charge of the black hole, where the heat capacity at constant chemical potential diverges thereby indicating the presence of a phase transition. Absence of any discontinuity in the $T-S$ curves eliminates the presence of any first-order transition. We then performed a detailed analysis of the two Ehrenfest's equations using numerical techniques reported in Tab.~\ref{tab2}. The results reveal that both of Ehrenfest's equations are satisfied and thus the phase transition is of second-order.

We obtained the thermodynamic curvature of the Weinhold, Ruppinier and NTG metrics for the charged black holes in dJT gravity and showed that they are exactly divergent at the  critical points; (i) at $T_c$ with the power-law $R_{NTG}\sim {(T  - T_{c} )} ^{-2}$, and (ii) at $T'_{c}$ one has $R_{W}\sim {(T  - T'_{c} )} ^{-2}$ and $R_{Ru}\sim {(T  - T'_{c}  )} ^{-2}$. It was shown that $R_{W}$ and $R_{Ru}$ curvatures have linear relationship with temperature at special value $\xi=2$ which is akin to that of heat capacity at constant chemical potential. Moreover, from Fig.~\ref{fig:6} it was investigated that the curvature $R_{NTG}$ behaves just like to the heat capacity at constant charge $C_{Q}$ for values of  $Q^2<Q_c^2$, $Q^2=Q_c^2$, and $Q^2>Q_c^2$ which they may be regarded as three different phases of black holes. The results are also consistent with the phase transitions studied in Ref.~\citep{Lu:2022tmt} from the page curve of this kind of black holes in dJT gravity.

As outlook, there are two concept that can be investigated for charged black holes in dJT gravity, one in the thermodynamic geometry and the other in the holographic complexity.
One of the topics that is studied in thermodynamic geometry is the extrinsic curvature, which is defined as $\mathcal{K}=\partial_{\mu}n^{\mu}$, where $n^{\mu} = (0,n_{Q})=\frac{1}{\sqrt{g_{Q \,Q}}}(0,1)$ is the vector perpendicular to the thermodynamic hypersurface \cite{Mansoori:2016jer}.
Moreover, Refs.~\citep{HosseiniMansoori:2020jrx,Rafiee:2021hyj} reported a phenomenon of universal behavior near critical points, wherein the critical amplitude of $R_Nt^2$ and $\mathcal{K}_Nt$ is $-\frac{1}{2}$ as $t\rightarrow0^+$, while $R_Nt^2\approx -\frac{1}{8}$ and $\mathcal{K}_Nt\approx-\frac{1}{4}$ in the limit $t\rightarrow0^-$ for four dimensional charged AdS black holes. Here, the terms $\mathcal{K}_{N}$ and $R_{N}$ denote normalized extrinsic and intrinsic curvature, respectively, and $t$ is defined as $t=T/T_c-1$, where $T_c$ signifies the critical temperature value. This critical universality bears merit and will be investigated in our future research endeavors.

In recent studies, holographic complexity \cite{Susskind:2014rva,Stanford:2014jda,Alishahiha:2015rta} has been explored in black holes coupled to axion fields \cite{Andrade2014,Yekta:2020wup,Babaei-Aghbolagh:2021ast}. For such black holes, it has been observed that the growth rate of holographic complexity in the complexity equals volume picture (CV conjecture) vanishes at zero temperature ($T=0$), while in the complexity equals action picture (CA conjecture) it was vanished at a minimum temperature ($T\neq 0$). This behavior corresponds to the divergence of the thermodynamic curvature for Ruppeiner geometry \cite{Babaei-Aghbolagh:2022xcy}. Also, there has been interesting studies on the complexity of JT gravity in \cite{Alishahiha:2018swh,Cai:2020wpc}. Therefore, it is compelling to investigate the behavior of holographic complexity for dJT models at critical and zero temperature ($T_c,T=0$) to determine if the growth rate of complexity will be of a specific value. These are fascinating questions that we will address in future.
\section{Acknowledgment}
We would like to thank Daniel Grumiller  and  Seyed Ali Hosseini Mansoori for useful discussions.


\bibliography{refs}
\end{document}